\documentclass[review]{elsarticle}

\usepackage{hyperref}
\usepackage{soul,xcolor}
\setstcolor{red}

\usepackage{multirow}
\usepackage{lscape}
\usepackage[toc,page]{appendix}
\usepackage{pifont}
\newcommand{\cmark}{\ding{51}}%
\newcommand{\xmark}{\ding{55}}%
\usepackage{rotating}
\usepackage{tablefootnote}
\usepackage{tabularx}
\usepackage[T1]{fontenc}
\usepackage[utf8]{inputenc}
\usepackage{csquotes} 
\usepackage[font=footnotesize,labelfont=bf]{caption}
\usepackage[font=footnotesize,labelfont=bf]{subcaption}
\usepackage{graphicx}
\usepackage{amsmath,amssymb,amsfonts}
\usepackage{algorithm,algorithmicx,algpseudocode}
\algrenewcommand\algorithmicrequire{\textbf{Input:}}
\algrenewcommand\algorithmicensure{\textbf{Output:}}
\algdef{SE}[DOWHILE]{Do}{doWhile}{\algorithmicdo}[1]{\algorithmicwhile\ #1}%
\usepackage{url}

\usepackage{hyperref}
\usepackage{array}

\usepackage{color, colortbl}

\definecolor{ashgrey}{rgb}{0.7, 0.75, 0.71}
\usepackage{geometry}

\usepackage{booktabs, ltablex}
\keepXColumns
\usepackage[version=4]{mhchem}
\usepackage{siunitx}

\usepackage{enumitem}        
\newlist{tabitem}{itemize}{1}
\setlist[tabitem]{nosep,     
                     topsep     = 0pt       ,
                     partopsep  = 0pt       ,
                     leftmargin = *         ,
                     label      = $\bullet$ ,
                     before     = \vspace{\baselineskip},
                     after      = \vspace{-\baselineskip}
                     }
\usepackage{pdflscape}

\usepackage{setspace}   
\onehalfspace           

\usepackage[multiple]{footmisc}
\newcolumntype{L}[1]{>{\raggedright\let\newline\\\arraybackslash\hspace{0pt}}m{#1}}
\newcolumntype{C}[1]{>{\centering\let\newline\\\arraybackslash\hspace{0pt}}m{#1}}
\newcolumntype{R}[1]{>{\raggedleft\let\newline\\\arraybackslash\hspace{0pt}}m{#1}}
\usepackage{tikz}
\usetikzlibrary{fadings}
\usetikzlibrary{arrows.meta}
\tikzset{%
  >={Latex[width=2mm,length=2mm]},
            base/.style = {rectangle,  draw=black,
                           minimum width=2.2cm, minimum height=1cm,
                           text centered, font=\sffamily},
  activityStarts/.style = {base, fill=blue!20, middle color=white},
         process/.style = {base, minimum width=2.2cm, fill=blue!20, middle color=white,
                           font=\ttfamily},
}

\usepackage{pdflscape,array,booktabs}
   \usepackage{lscape}
    \newlist{tableitems}{itemize}{1}
    \usepackage{mathabx}
    \setlist[tableitems]{nosep,
                         topsep=0pt,
                         partopsep=0pt,
                         leftmargin=1em,
                         label=$\sqbullet$
    }

\begin{document}

\title{Orchestrating Product Provenance Story: When IOTA Ecosystem Meets The Electronics Supply Chain Space}

\author[1]{Sabah~Suhail \corref{a}}
\cortext[a]{Corresponding author}
\ead{sabah@khu.ac.kr}

\author[2]{Rasheed Hussain}
\ead{r.hussain@innopolis.ru}

\author[3]{Abid~Khan}

\ead{abidkhan@comsats.edu.pk}

\author[1]{Choong~Seon~Hong}
\ead{cshong@khu.ac.kr}

\address [1]{Department of Computer Science and Engineering, Kyung Hee University, South Korea.}
\address[2]{Networks and Blockchain Lab, Innopolis University, Russia.}
\address[3]{Department of Computer Science, Aberystwyth University, United Kingdom.}

\begin{frontmatter}
\begin{abstract}
\textit{\enquote{Trustworthy data}} is the fuel for ensuring transparent traceability, precise decision-making, and cogent coordination in the Supply Chain (SC) space. However, the disparate data silos act as a trade barrier in orchestrating the provenance of the product lifecycle; starting from the raw materials to end products available for customers. Besides product traceability, the legacy SCs face several other problems including, data validation, data accessibility, security, and privacy issues. \emph{Blockchain} as one of the advanced \emph{Distributed Ledger Technology} (DLT) addresses these challenges by linking fragmented and siloed SC events in an immutable audit trail. Nevertheless, the challenging constraints of blockchain, including, but not limited to, scalability, inability to access off-line data, vulnerability to quantum attacks, and high transaction fees necessitate a new mechanism to overcome the inefficiencies of the current blockchain design. In this regard, \emph{IOTA} (the third generation of DLT) leverages a Directed Acyclic Graph (DAG)-based data structure in contrast to linear data structure of blockchain to bridge such gaps and facilitate a scalable, quantum-resistant, and miner-free solution for the Internet of Things (IoT) in addition to the significant features provided by the blockchain. Realizing the crucial requirement of traceability and considering the limitations of blockchain in SC, we propose a provenance-enabled framework for the Electronics Supply Chain (ESC) through a permissioned IOTA ledger. To identify operational disruptions or counterfeiting issues, we devise a transparent \emph{product ledger} constructed based on trade event details along with timestamped sensory data to track and trace the complete product journey at each intermediary step throughout the SC processes. We further exploit the \emph{Masked Authenticated Messaging} (MAM) protocol provided by IOTA that allows the SC players to procure distributed information while keeping confidential trade flows, ensuring restrictions on data retrieval, and facilitating the integration of fine-grained or coarse-grained data accessibility.
Our experimental results show that the time required to construct secure provenance data aggregated from multiple SC entities takes on average 3 seconds for a local node and 4 seconds for a remote node, which is justifiable.
Furthermore, to estimate the energy consumption by the IoT platform, we use Raspberry Pi 3B to ensure that the energy consumption by the proposed scheme is admissible by resource-constrained devices. 
\end{abstract}


\end{frontmatter}


\section{Introduction}\label{introduction}
Electronics Supply Chain (ESC) revolves around an intricate and intensive process during which raw materials or natural resources are transformed into circuit boards and electronic components, integrated and assembled into end products, and ultimately made available to the customers. 
Such a complex product evolution journey involving collaboration among multiple Supply Chain (SC) participating entities, each performing different operations on a product (or its parts), may raise several questions. For instance, how to identify the granular details of the underlying processes such as who, when, what, where, and how the product was derived. To answer these questions, SCs need a \emph{track and trace} mechanism called \emph{provenance} to construct a complete lineage of data about products' origin, production, modification, and custody process~\cite{montecchi2019s}. Provenance in SC can enable the enterprises to choreograph their demand-supply circle, perform risk assessment, maximize revenues, investigate reasons for product recalls, and forecast their future goals. Furthermore, provenance ensures the integrity of data during data debugging, reconciliation, replication, decision making, performance tuning, auditing, and forensic analysis~\cite{cheney2009provenance,zafar2017trustworthy,suhail2018data,suhail2016introducing, suhail2020provenance}.
However, procuring product provenance data is an exhaustive task which gives rise to several other challenging issues concerning the collection, distribution, accessibility, and security of data. For instance, (i) how to collate provenance data from disparate data silos, complex data aggregation processes, and on-premise operational practices and procedures, (ii) how to assure integrity, reliability, and resiliency of data, (iii) how to ensure the distributed data accessibility and availability to legitimate participating entities, and many more. 

Due to the unavailability of a platform that can provide {\it one-size-fits-all} solution to orchestrate product provenance story, it is hard to differentiate between reliable and counterfeit products. To this end, the proliferation of counterfeit products deteriorates consumer trust and also causes reputational damage to the company's image. For instance, defense system manufacturers face difficulty in detecting counterfeit items, as counterfeiters attempt to imitate materials, part numbers, and serial numbers to simulate authentic parts~\cite{stradley2006electronic}.
Similarly, Integrated Circuits (ICs) counterfeiting has been observed in many industrial sectors, including computers, telecommunications, and automotive electronics~\cite{guin2014counterfeit}.
For example, in 2018 Orange County electronics distributor was charged with selling counterfeit integrated circuits for military and commercial use \cite{Orange}.
To mitigate the risk effectively across the SC, \textit{atomistic} sources of risk are required that involves scrutinizing a restricted part of the SC and is suitable for low-value and less complex components. Alternatively, \textit{holistic} sources of risk are required that involves a comprehensive analysis of the SC and is preferable for high-value and complex components~\cite{svensson2004key}. In both of these cases, contingency planning is required to identify the root cause of operational disruption and to identify the fraudulent middleman.

The inception of Distributed Ledger Technology (DLT) solves SC 
challenges by facilitating distributed, immutable, transparent, and fault-tolerant data aggregation across multiple entities (both physical and digital)~\cite{babich2019distributed}. In this regard, the blockchain-based architecture is used as a potential solution to fulfill the digital SC requirements in a plethora of SC use-cases (as discussed in Table \ref{tab:BCinSC}). 
Table \ref{tab:BCinSC} outlines the current research efforts that use blockchain-based solutions for SC use-cases while considering their security aspects.


However, the current blockchain solutions lack many striking features such as scalability, offline data accessibility, fee-less transactions, and quantum-immunity that are among the desirable features in digital SC. To adequately address these limitations, IOTA~\cite{popov2017iota} brings a transformation in the third generation of DLT. Following a scalable and quantum-immune approach, it securely accelerates tracking and tracing of multiple trade events in the SC even in the offline mode, and consequently enhance provenance data construction to identify counterfeit products.

\begin{figure}[!ht]
\centerline{\includegraphics[width=3.0in]{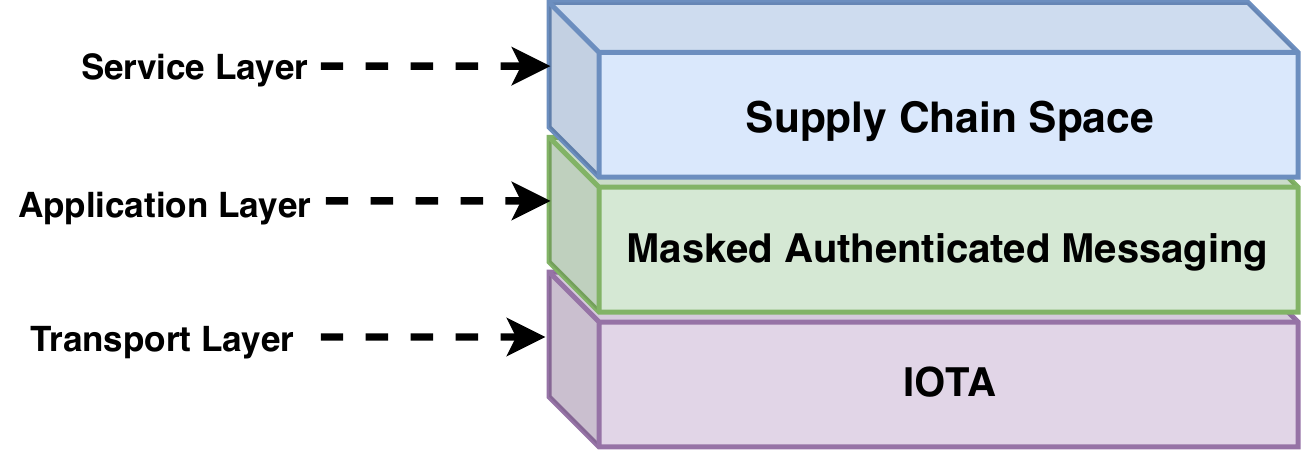}}
\caption{ Abstract-level overview of the proposed framework.}
\label{fig:sysmodel}
\end{figure}

In this paper, we propose an IOTA-based framework for supporting provenance in the ESC. By integrating the Masked Authenticated Messaging (MAM) protocol on top of IOTA (as shown in Fig.~\ref{fig:sysmodel}), our proposed framework provides transparent traceability of data throughout the SC ensuring trustworthy and quality data.   

The main contributions of this paper are summarized as follows.
\begin{itemize}
     \item We present a technological evolution of the SC management system starting from legacy systems to the state-of-art DLT. We also investigate the significance of integrating provenance in the ESC and address the research gaps and highlight the key factors for adopting IOTA in the SC in comparison to existing blockchain-based solutions. 
     
    \item For product traceability, we propose an IOTA-based provenance framework that encapsulates the diverse \textit{product story} as \textit{provenance data} at each intermediary process in the ESC. Such a strategy solves the product counterfeiting issue in addition to the problem of fragmented and asymmetric information. To address the security issues, the MAM channel is leveraged to ensure confidential trade flow among competitors, to preserve data integrity, and to provide fine-grained data access to the trusted SC players only.
    
    \item Finally, we evaluate and show that the proposed scheme in terms of attaching SC data to the IOTA ledger and constructing provenance data by fetching data for varying payload sizes is admissible for the ESC. Furthermore, we develop a proof-of-concept for the proposed scheme on the Raspberry Pi 3B hardware platform to mimic the IoT-integrated ESC and analyze the measured average time and energy consumption incurred during attaching and fetching provenance data from the IOTA tangle which we find justifiable for resource-constrained IoT devices deployed in the SC.
    \end{itemize}
The rest of the paper is organized as follows. Section \ref{relatedwork} surveys related work. Section \ref{background} provides an overview of IOTA, SC, and discusses the significance of integrating IOTA to support provenance in SC. Section \ref{SystemModel} introduces the system model and Section \ref{scheme} describes the proposed IOTA framework for the ESC. Section \ref{simulation} presents the simulation results and discusses the security analysis of the proposed approach. Finally, Section \ref{conclusion} concludes the paper with an outlook on future work.

\begin{landscape}
\begin{table}[!ht]

 \caption{Current research efforts pertaining to blockchain-based solutions for SC use-cases and their security aspects.} \label{tab:BCinSC}
\begin{tabular}{>{\raggedright}p{4.0cm}lccccccccccccc}
\hline
 & \multicolumn{12}{c}{\textbf{SC Technical Challenges}}       \\
 \cline{3-14}
\textbf{SC Category}  & \textbf{Scheme}  & D & C & I & A & S & AC & Audit & T & OFD & IoT/IIoT & BCID & QI\\
 \midrule

Food/Agriculture & \cite{tian2017supply} & \cmark  & \cmark & \cmark & \cmark & $\circleddash$ & \xmark & \xmark & \xmark& \xmark & $\circleddash$ & \xmark & \xmark \\

& \cite{biswas2017blockchain}& \cmark &  \cmark& \cmark & \cmark & \xmark & \xmark & \xmark & \xmark & \xmark & \xmark & \xmark & \xmark \\

& \cite{malik2018productchain} & \cmark & \cmark& \cmark& \cmark & \cmark & \cmark & \cmark & \xmark & \xmark & \cmark & \cmark & \xmark \\

& \cite{tsang2019blockchain} &\cmark& \cmark & \cmark & \cmark & $\circleddash$ & \xmark & \xmark & \xmark & \xmark & \cmark & \cmark & \xmark \\

& \cite{caro2018blockchain}&\cmark& \cmark & \cmark & \cmark & \xmark & \xmark & \xmark & \xmark & \xmark & $\circleddash$ &\xmark & \xmark \\

&\cite{tian2016agri} &\cmark& \cmark & \cmark & \cmark & \xmark & \xmark & \xmark & \xmark & \xmark & $\circleddash$ & \xmark & \xmark \\

\hline
Pharmaceutical/Healthcare &\cite{raj2019anticounterfeiting}& \cmark & \cmark &  \cmark &  \cmark &  \cmark &  \xmark &  \xmark & \xmark & \xmark & \xmark & $\circleddash$ & \xmark\\

&\cite{sylim2018blockchain}&  \cmark &  \cmark &  \cmark &  \cmark &  \xmark & $\circleddash$ & $\circleddash$ &  \xmark &  \xmark &  \xmark &  \xmark & \xmark\\

&\cite{bocek2017blockchains}&  \cmark &  \cmark &  \cmark &  \cmark & \xmark &  \xmark &  $\circleddash$ &  \xmark &  \xmark &  \cmark &  \cmark & \xmark\\

\hline
Electronics & \cite{cui2019blockchain} &  \cmark &  \cmark &  \cmark &  \cmark & $\circleddash$ & \cmark & \cmark & $\circleddash$ & \xmark & \xmark & \cmark & \xmark\\

& \cite{xu2019electronics} &  \cmark &  \cmark &  \cmark &  \cmark &  $\circleddash$ &$\circleddash$& \xmark & $\circleddash$ &  \xmark & \cmark & \xmark & \xmark\\

\hline
Generic  &\cite{sidorov2019ultralightweight}&  \cmark &  \cmark &  \cmark &  \cmark &  \xmark &  \cmark  &  \xmark &  $\circleddash$ &  \xmark &  \xmark &  \cmark & \xmark  \\

&\cite{westerkamp2019tracing} & \cmark &  \cmark &  \cmark &  \cmark & \cmark & \cmark & $\circleddash$ & \xmark & \xmark & \xmark & \cmark & \xmark\\

&\cite{wu2017distributed}& \cmark &  \cmark &  \cmark &  \cmark & \xmark & $\circleddash$ & \xmark & \xmark  & \xmark  & \cmark  & \cmark & \xmark \\

&\cite{tunstad2019hyperprov}& \cmark &  \cmark &  \cmark &  \cmark & \cmark & \cmark & \cmark & \xmark  & \xmark  & \cmark  & \cmark & \xmark  \\

\hline
\bottomrule
\multicolumn{12}{l}{$\circleddash$=discussed but details are not provided.
}                                \\
\end{tabular}\\
D=Distributed, C=Confidentiality, I=Integrity,  A=Availability, S=Scalability,  AC=Access Control,\\ Audit=Auditability, T=Trust, OFD=Off-line data, BCID=Blockchain Implementation Details, QI=Quantum Immune.

\end{table}
\end{landscape}

\section{Related Works} \label{relatedwork}
In this section, we discuss the existing solutions for SC. 

\subsection{Legacy Systems: Enterprise Software Solutions for SC}

In literature, many works analyzed the limitations of ERP systems. For example, \cite{shehab2004enterprise} discussed the reasons for ERP dissatisfaction including, huge storage and networking requirements, training overheads, business process re-engineering, and software-related customization tasks. \cite{yen2002synergic} identified disadvantages of ERP such as privacy concerns, high cost of ERP software incurred during the implementation, maintenance, upgrading, and time-consuming customization, to name a few. Similarly, in \cite{soh2000cultural}, the authors discussed ERP system misfits (in terms of data, functional, and output) that arise from incompatibilities between ERP packages and organizational requirements. 
In addition to the aforementioned drawbacks, handling transactions in ERP systems are complex and rampant because each organization maintains its own ledger. Due to this fact, information discontinuity and desynchronization as a weak link may increase the possibility of human error or fraud whereas reliance on a middleman for validation leads to inefficiencies.

\subsection{Cloud-Based Solution for SC}
In comparison to ERP systems, cloud-based solutions, for instance, Software as a Service--SaaS, are plugged into SC to provide several benefits including fault tolerance, high availability, scalability, flexibility, reduced overhead for users, reduced cost of ownership, on-demand services, etc. \cite{tiwari2013analysis}. 
However, the adoption of cloud-based solutions may create new risks to the SC. For instance, trusting a third-party platform to store confidential information related to SC system, applications, and customer data can raise security and privacy issues \cite{truong2010cloud, durowoju2011impact}.
Similarly, storing data or sensor values in a secure and confidential manner is still under question in cloud-based solutions. Hence, introducing cloud-based solutions raise significant concerns about privacy, security, data integrity, intellectual property management, audit trails, compatibility, and reliability \cite{smith2009computing, hayes2008cloud}.
Moreover, in case of a cloud service outage, all SC operations will have a catastrophic effect on the data availability and ultimately deteriorate the overall performance of the system \cite{kim2009adoption, duan2013benefits}. 

\subsection{DLT-Based Solution for SC}
The imperative transition to adopt DLT in SC received significant attention in recent years as a solution to many challenges stemmed from the existing legacy system and privacy issues due to the multi-party access \cite{usecaseAnalysis}. In this regard, the nascent technology of blockchain allows companies to record every event within a SC on a distributed ledger that is shared among all participants, not owned by anyone, solve important glitches in traceability, and record events in a secure, immutable, and irrevocable way. 

Blockchain has a very constructive role in SC from different perspectives. For instance, blockchain can provide design decisions and solve technical challenges (as shown in Fig. \ref{fig:features}) faced by SC. Furthermore, blockchain-based architecture for SC that leverages public or private blockchain, can use different platforms such as Ethereum, Hyperledger Fabric, and Ripple. 
Many promising blockchain-enabled solutions have been proposed in literature where blockchain is leveraged in SC across different industries, for instance, food, agriculture, pharmaceutical, electronics SCs, to name a few. In Table~\ref{tab:BCinSC}, we summarized technical blockchain-based solutions for various SC categories (application-specific or generalized solutions) and associated shortcomings either in the context of the proposed scheme or the blockchain in particular. In Industry 4.0, blockchain can automate processes among IIoT, cyber-physical systems, and supply partners~\cite{ghobakhloo2018future}. For example,~\cite{li2018toward} discusses the integration of blockchain into the manufacturing industry for data integrity and resilience. Similarly, in \cite{vatankhah2019blockchain}, the authors propose a block-chain based platform for small and medium manufacturing enterprises (SMEs) to solve issues such as security, scalability, and big data problems.

In the following, we discuss non-technical works in literature that focus on the technical and non-technical challenges of blockchain-based deployed systems in SCs, for example, in~\cite{kamilaris2019rise}, the authors highlight technical, educational, and regulatory challenges and barriers in agriculture and food SCs. 
In \cite{khezr2019blockchain}, the authors discuss open research challenges in blockchain-based use-cases, including the Internet of Medical Things (IoMT), healthcare data management, and SC management. Furthermore, the authors of~\cite{clauson2018leveraging} also provide an overview of the challenges associated with blockchain adoption and deployment for the health SC with a focus on pharmaceutical, Internet of Healthy Things (IoHT), and public health.  In \cite{lee2017blockchain}, the authors discuss the factors that bring a positive impact on blockchain-based ESC. Similarly, many other use cases are discussed in~\cite{kshetri20181, reyna2018blockchain, hughes2019beyond, morkunas2019blockchain}. 

Recently, various worldwide enterprises have played a significant role in providing blockchain-based platforms for supporting friction-less traceability and transparency in SC, for instance, IBM's blockchain framework~\cite{IBMBC} has been adopted by Walmart, Nestle, Unilever, and other players in the global Food Supply Chain (FSC)~\cite{IBMblockchain}. Other notable blockchain-enabled SC frameworks include Hyperledger \cite{Hyperledger}, skuchain \cite{skuchain}, Provenance \cite{Provenance}, Blockverify \cite{blockverify}, Multichain \cite{multichain}, etc. However, the proprietary and private blockchain-based solutions are unable to address the specific requirements or needs in the public domain and portray blockchain as a black-box.

\subsection{Limitations of Blockchain in SC} \label{limitations}
Most of the blockchain-based solutions adopted for SC theoretically cover advantages, potential challenges, and future directions \cite{saberi2019blockchain, hackius2017blockchain, wang2019making, wang2019understanding, francisco2018supply}. However, the underlying constraints of blockchain are overlooked by the current solutions. Among other constraints, quantum-resistance and scalability are noteworthy. Ongoing efforts to address these potential issues are underway, for instance, to meet the challenging requirement of quantum future, some of the emerging blockchain
solutions that already support post-quantum techniques are
Quantum Resistant Ledger (QRL)~\cite{QRL}, Corda~\cite{brown2016corda}, Quantum-secured blockchain~\cite{kiktenko2018quantum}, and DL-for-IoT~\cite{shahid2020post}, etc.
Solutions such as sharding and off-chain are expected to solve the scalability problem of the blockchain. However, these solutions have their own drawbacks. 
For instance, sharding requires to synchronize the running of operations among different processes on different shards. Furthermore, the overheating of a targeted single shard due to many cross-shard transactions is another problem that requires the ranking of these transactions to prevent overloading of block producers on the target shard.
Similarly, off-chain solutions suffer from the following limitations: (i) 
introduce additional layers of complexity as the protocols are built on the top of the blockchain, (ii) may face objection by government and business communities due to their censorship-resistant nature.
Directed Acyclic Graph (DAG)-based blockchain design is another effort to overcome the scalability issue caused by the sequential chain-based design of the traditional blockchain~\cite{babich2019blockchain}.
The authors of~\cite{benvcic2018distributed, pervez2018comparative, raderblockchain} provide a comparative analysis of DAG-based blockchain schemes. 

Many of current research works are assuming, though some with skepticism, that blockchain technology will be adopted in the SC industry.
For instance, some of their concerns are as follows: considering the connection between physical and digital world, how to ensure the reliability of data from SC entities and sensors~\cite{wust2018you}, security concerns due to quantum computing and latency issues with the increasing number of nodes in the network~\cite{occam, prewett2019blockchain}, lack of privacy and Garbage In Garbage Out (GIGO) problem~\cite{babich2019distributed}, lack of information leading to existence of gray markets \cite{babich2019blockchain} and decision paralysis due to information overload, etc. 
But paradoxically on the other side, solutions such as \cite{montecchi2019s, wang2019making, roeck2019distributed, saberi2019blockchain, wang2019understanding, azzi2019power, sternberg2018chains} focus on the significance of using blockchain in SC.
Overall, most of the proposed schemes (discussed in Table \ref{tab:BCinSC}) failed to address the potential current problem (such as scalability) and future issues (such as quantum-resistance) of blockchain. Moreover, other technical requirements of SC such as accessibility and auditability based on roles and access levels, are also overlooked. 

The common denominator among DLTs is their reliance on a distributed, decentralized peer-to-peer network, and consensus mechanism. However, DLTs vary substantially in terms of the underlying data structure, fault tolerance, and consensus approaches~\cite{el2018review}. In addition to blockchain and its different flavors, other well-known DLTs are tangle, hashgraph~\cite{baird2016swirlds}, sidechain, and holochains~\cite{harris2018holochain}. In \cite{el2018review}, the authors provide a comparative analysis of DLTs, whereas, in \cite{pervez2018comparative}, the authors compare classical blockchain with DAG-based blockchain.
Considering the primary challenge, i.e., scalability, faced by blockchain-based solutions in SCs, we consider a DAG-based DLT, i.e., IOTA. In comparison to other DLTs, IOTA exhibits quantum immune nature, provides off-line data accessibility, and supports fee-less microtransactions that are important factors for future SCs. Some of the striking features of IOTA DLT are discussed in Section~\ref{WHY}.

\subsection{Our Research Contributions}
In the presence of the existing research, we highlight the main contributions of our work. This research is aimed to highlight the current research gaps in SCs and propose a state-of-the-art approach to resolve them.
Though the existing blockchain-based solutions for SC has solved the primitive problems associated with disjoint data fragments, third party dependency, data security, and many other problems related to legacy systems. Nevertheless, there are still a few issues with the existing solutions that need to be addressed. In this regard, our contributions include the following key factors that are required to incorporate in a DLT-enabled SC. 

Firstly, \textit{why the transition from the mainstream blockchain to IOTA is required?} We adopt IOTA DLT upon realizing the overlooked constraints of blockchain. For instance, scalability issues particularly in case of growing participating entities; accessing data from freights in remote areas or off-line mode; dealing with transaction fees, and finally reliance on the security of current cryptographic primitives keeping in view the not-so-far arrival of quantum computers. The details about such constraints are discussed in Section~\ref{WHY}.
Secondly, \textit{how to create transparency towards the consumers?} To win consumer trust, it is important to give them a sheer picture of the product journey. We devise a mechanism that involves reconstructing a trustworthy product provenance story, thus enabling customers to drive a real change.
Thirdly, \textit{how to define customized data access control rights?} We use the MAM protocol to provide fine-grained data access privileges to facilitate the trade secrets of participating entities. 
Fourthly, \textit{how to identify counterfeit products?} We construct provenance data such that it includes complete information to identify the illegitimate or defective item.


For illustrative purposes, we consider the example of mobile phones in ESC. While keeping the underlying framework same, the proposed model is suitable for any commodity in ESC. Furthermore, it can be customized to other non-electronics SCs (for example, food-agriculture, pharmaceuticals) keeping in view the diverse requirements driven by their specific business needs and additional information (e.g., expiry dates or any other precautionary measures). For instance, food-agriculture, pharmaceuticals, or any other cold chain differs from ESC as they are subject to sensitive temperature and environmental conditions necessary to maintain the quality of perishable items in terms of temperature, humidity, etc. Similarly, ESC differs from other SCs based on quality testing, such as expiry period in case of cold chains are completely different from the warranty period determined through failure testing/product life testing of electronic components. Other differences include the packaging and assembling of components at various stages. 
Such requirements of cold SCs can be facilitated through our proposed model by continuous monitoring and reporting of the sensor data at frequent intervals to ensure the quality of products while allowing the integration of any optional information. Therefore, by tweaking the parameters based on the details of the underlying SC case, the proposed framework can be applied to any other SC.

\section{IOTA in SC: An Overview} \label{background}
In this section, we provide a quick overview of the key terms in IOTA and SC. We also emphasize on using IOTA DLT in the SC. 
\subsection{IOTA}
\begin{figure}[!ht]
\centerline{\includegraphics[width=4.5in]{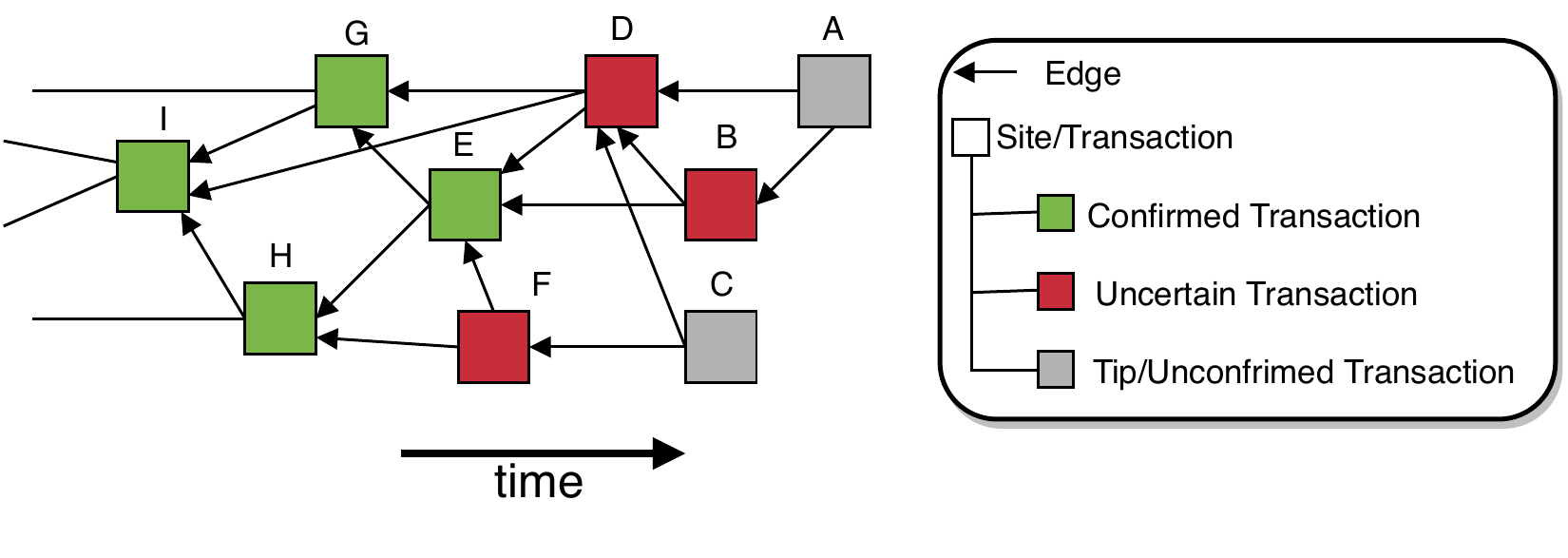}}
\caption{Tangle graph illustrating transactions and edge set connecting transactions.}
\label{fig:tangle}
\end{figure}

IOTA is a public, permissionless, and a distributed ledger that leverages directed acyclic graph (DAG) data structure termed as \textit{tangle} for storing interlinked but individual transactions exchanged among peers \cite{popov2017iota}. Fig. \ref{fig:tangle} shows a tangle graph where each square-block represents a \textit{transaction/site} which is propagated by a \textit{node}. Every new transaction attached to the tangle graph forms an edge set. To create a transaction, a node (i) creates and signs a transaction with its private key, (ii) use the Markov Chain Monte Carlo (MCMC) algorithm \cite{gilks1995markov} to choose and validate two other non-conflicting unconfirmed transactions (tips), and (iii) solve a cryptographic puzzle (known as Hashcash~\cite{back2002hashcash}) to perform Proof of Work (PoW) for preventing Sybil attack. Transaction status can be categorized as confirmed transactions (green nodes), uncertain transactions (red nodes), and unconfirmed transactions or tips (grey nodes) as shown in Fig. \ref{fig:tangle}. The revolutionary features of IOTA including scalability, decentralization, zero transaction fee, speedy microtransactions, off-line capability, and quantum security enables it to gain ground not only in Machine-to-Machine (M2M) economy but also in application areas encompassing IIoT.

\subsection{Supply Chain System}
\begin{figure}[!ht]
\centerline{\includegraphics[width=2.5in]{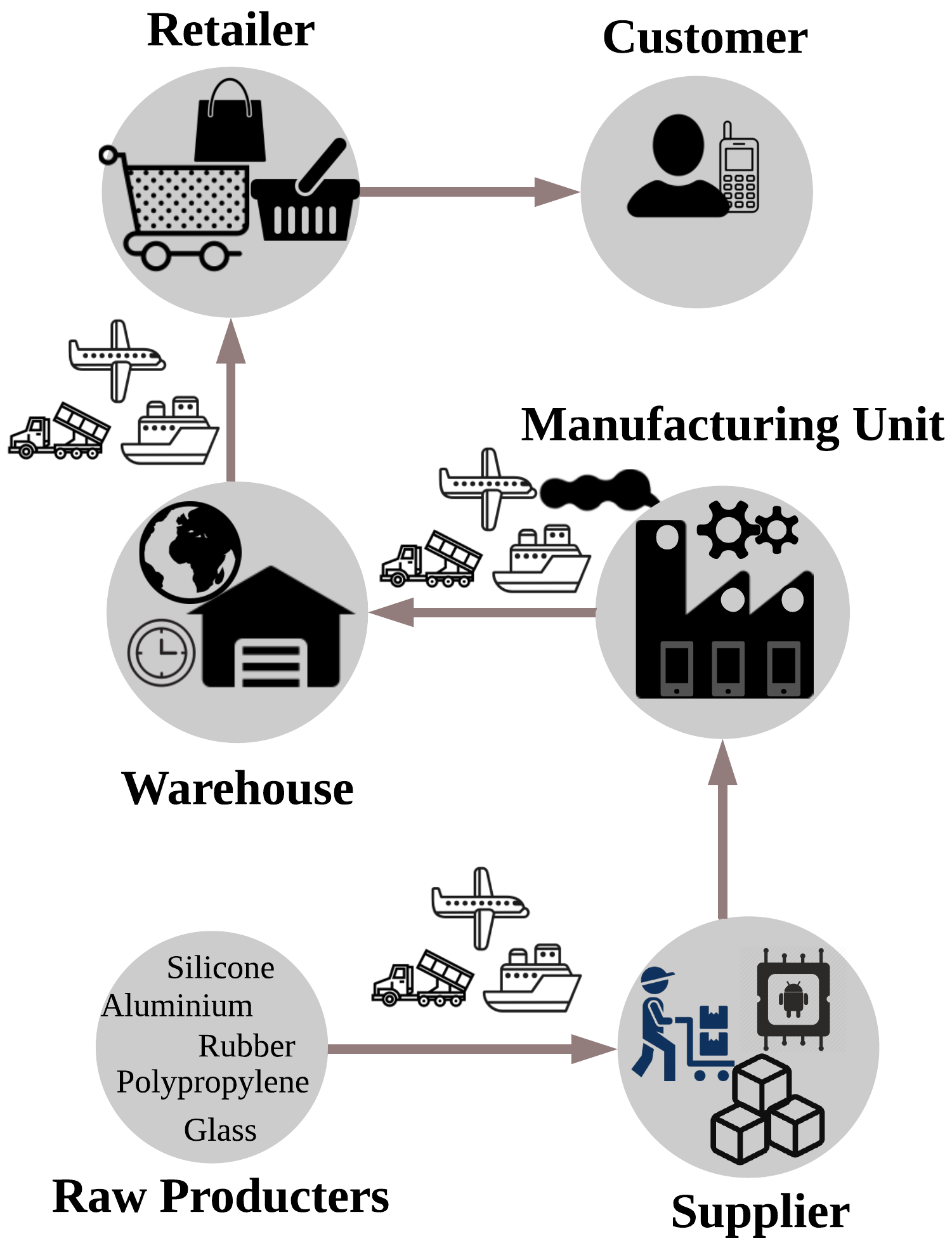}}
\caption{Electronic supply chain showing participating players involved in supply chain processes.}
\label{fig:supplychain}
\end{figure}

Supply chain
encompasses coordination and collaboration among channel partners (suppliers, intermediaries, third-party service providers, and customers) for planning and managing of upstream and downstream process-based activities such as the transformation of natural resources/raw materials, sourcing, procurement, production, conversion, and logistics~\cite{kozlenkova2015role}. Fig.~\ref{fig:supplychain} shows the primary entities involved in the production of mobile phones in the ESC.

\subsection{Motivation: Integrating Provenance Support in the Electronics Supply Chain Through IOTA} \label{IoTSC}
SCs empower participants for collaborative commerce in a global value chain. A product (for instance, mobile phone) journey from \emph{sand to hand}, comprise of numerous chained phases during which components are produced, sourced, refined, integrated, and assembled by multiple entities ubiquitously. Nevertheless, numerous friction points thwart SCs from accomplishing their maximum potential, for instance, opaque mechanics of global commerce, complexity (upstream and downstream), manual processes, and divergent standards. Furthermore, SCs are held back by imperfect and asymmetric information as huge volumes of veracious data are inaccessible to the SC players during cross-border trading resulting in communication gaps, increased costs, forfeited data, erroneous data aggregation, scattered information, market failures, or absence of markets at all. Therefore, it can be concluded that a sustainable SC stipulates two pivotal features to be incorporated: (i) \textit{Product story} (ii) \textit{Orchestrating episodes of product story}.

\subsubsection{Provenance: Product Story}
 To trace the audit trail of data, provenance plays a significant role in constructing a "product story" throughout the SC. The product story (or product traceability) enables the seller-buyer pair to trace the product from its inventory procurement process to its point of sale and hence provides an efficient way for tackling counterfeits or determine liability in the event of faulty records. For instance, upon scanning the Quick Response (QR) code, the consumer can look at the product story.

\subsubsection{IOTA: Orchestrating Episodes of Product Story }
Relying solely on provenance to notch trustworthy data is not sufficient, therefore, to efficiently and proactively systematize the product story in terms of provenance data, we collocate the product story episodes in a tamper-proof product ledger perpetuated by IOTA. IOTA is a promising technology that could play an indispensable role in providing salient features desirable for SC including decentralized, distributed, and interoperable data stream to consortium members. Furthermore, it provides data auditability to identify accountable actors causing data contamination, reasonable confidentiality and privacy of the trade flows, facilitating access control on immutable and trustworthy data. In essence, coalescing real-time provenance data through resource-constrained sensors (such as track and trace of precise location information, transfer of custody, monitoring environmental conditions or temperature fluctuations during storing and shipping product through GPS, RFID tags, temperature sensors, humidity sensors, etc.) that can drive decision-making power during risk mitigation forecasting.

\subsubsection{Striking Features of IOTA} \label{WHY}
IOTA leads to
a paradigm shift in the third generation of DLTs by providing a solution to the challenging constraints encountered by the blockchain. In essence, the IOTA platform breaks out the \textit{blockchain trilemma} as it exhibits the following salient features essential to SC in contrast to the blockchain.
\begin{enumerate}
\item \textbf{Scalability:}
In the SC space, the transaction load is expected to increase based on actors and activities in the network over time. Traditional blockchain runs into scaling snag as the transactions and the validation of said transactions are siloed and often at odds with each other. A permissioned blockchain has the potential to scale to a few hundred nodes \cite{li2017towards}, which may affect the latency and throughput of the network. For instance, the processed transactions
per second (TPS) of the mainstream blockchain platforms like Bitcoin (7 TPS) and
Ethereum (15 TPS) is very limited because the blocks cannot be created simultaneously due to a single linear chain of blocks~\cite{fan2019performance}. To address the problem of scalability, \cite{li2017towards,danezis2015centrally,gencer2017short} used the concept of sharding single blockchain ledger i.e., processing of transactions is distributed across multiple nodes in parallel. However, the block-less design of IOTA scales horizontally can theoretically improve transaction efficiency and claims to resolve the problem of scalability without any need to take care of global and local ledgers.

\item \textbf{IOTA for IoT:}
Cost-effective miniaturized things or objects with storage, computational, and networking capabilities are providing promising solutions to optimize myriad events of track-and-trace in the supply cycle. Such synergistic integration of IoT and SC results in the cyber-physical system concept of \textit{Industry 4.0}~\cite{frank2019industry}. However, IoT-linked security crises may breach the system security making it vulnerable to attacks, for instance, NotPetya, Stuxnet, PyPi, ShadowPad, and many other SC attacks. To cope with such situations, IOTA provides a high end-to-end security level called \textit{Masked Authenticated Messaging} (MAM) protocol that facilitates encrypted data communication and securely anchors it to the tangle in a quantum-proof fashion. Moreover, through the MAM channel, access privileges can be tailored, thereby making it a perfect platform to build tomorrow’s trade facilitation systems in SC.

\item \textbf{Off-line transaction:}
IOTA supports the off-line capability for handling scenarios such as information retrieval from remote areas and sensors on a container of a freighter ship that loses connectivity during ocean transportation. This mechanism is called "partitioning" during which a tangle in IOTA can get branch off by creating an off-line cluster and back into the network upon Internet connectivity, thereby securing the local transactions.

\item \textbf{Speedy and zero-fee transactions without miners:}
Blockchain can be used for payments in many notable application areas. However, PoW based blockchains~\cite{nakamoto2019bitcoin} require transaction fees due to the presence of miners and are not suitable for micropayments. Similarly, private, permission-based blockchains are lightweight~\cite{dorri2016blockchain} but lack important
properties for PoW based blockchains, such as censorship resistance~\cite{elsts2018distributed}. Unlike blockchains, IOTA is based on the concept of \textit{users equal to validators}. IOTA is free from miners i.e., all participants need to contribute their computation power to maintain the network and hence does not charge any transaction fee. Furthermore, by enabling transactions to be validated in parallel, IOTA has the highest TPS among other distributed ledger. 

\begin{figure}[!ht]
\centerline{\includegraphics[width=2.5in]{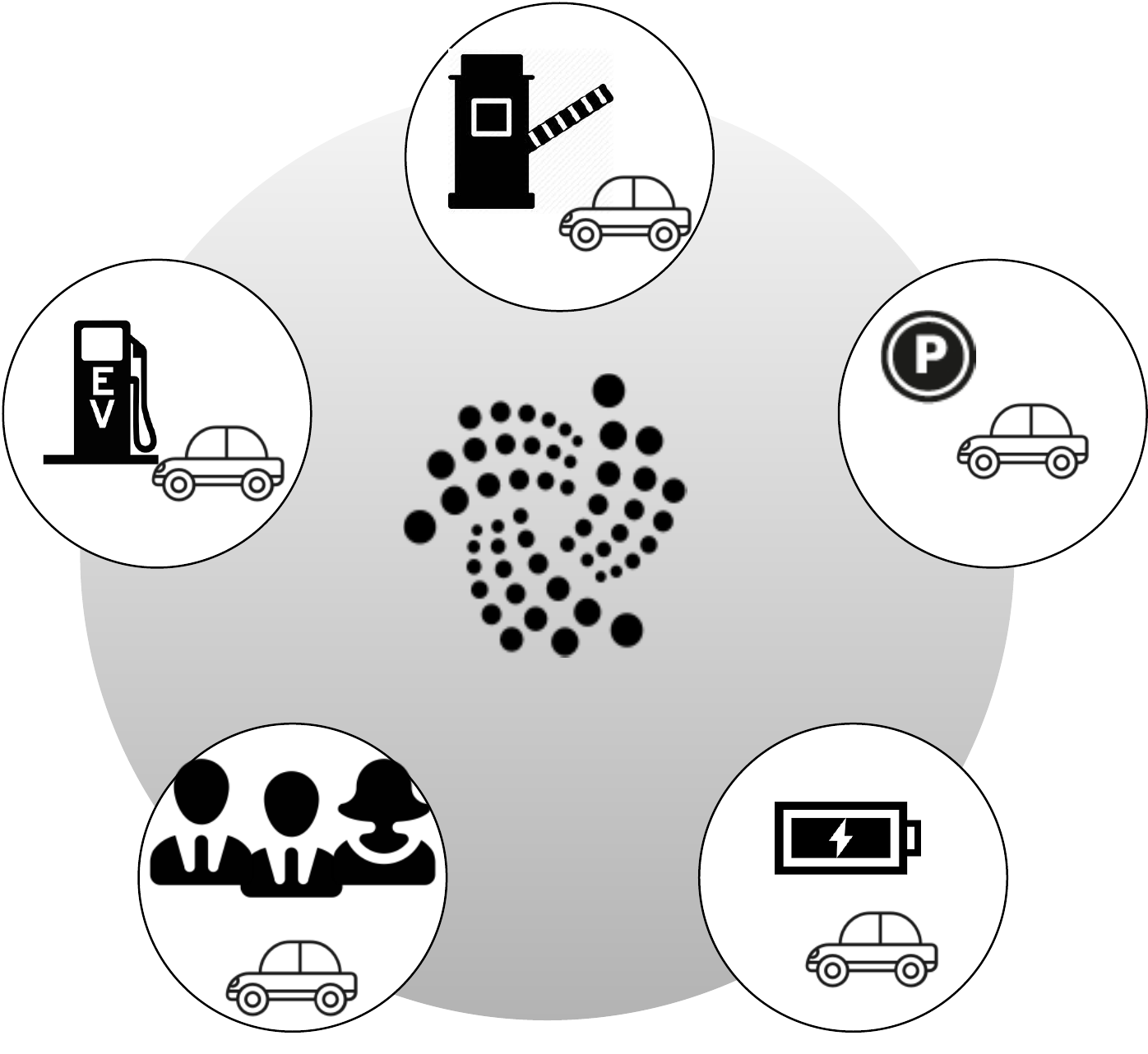}}
\caption{IOTA use case: machines with digital wallets.}
\label{fig:EV}
\end{figure}
\item \textbf{Machines with digital wallets:}
Though blockchain has also enabled M2M communications, for instance, \cite{sikorski2017blockchain} presents an example of the chemical industry where blockchain is employed to establish an M2M electricity market. However, the chained-based blockchains have certain limitations as discussed in Section \ref{limitations}. 

IOTA has also powered resource-constrained devices to participate in the M2M world. Being a powerful catalyst in the M2M economy, it allows machines to interact seamlessly and instantly transact with each other through its bi-directional off-tangle payment channel called \textit{Flash Channel} allowing massive volumes of nano and microtransactions~\cite{freiberg2017instant}. One such electric autonomous vehicle example scenario is suggested in~\cite{strugar2018m2m}. Thus, IOTA can enable the automotive to pay autonomously for services including parking, charging batteries, tolls taxes, usage-based insurances, and many more~\cite{iotaAuto} (as shown in Fig. \ref{fig:EV}). In September 2018, the Netherlands successfully completes testing for electric car charging station using IOTA~\cite{IotaEcar}. Another practical application of a digital twin with IOTA is CarPass~\cite{carpass} for vehicle telematics data  (e.g. mileage, trips, environmental, maintenance data)~\cite{digitalTwin}. The data marketplace is also an interesting application that simulates the connected devices running IOTA protocol to share secure and valuable real-time data~\cite{IOTADataMarketplace}.

\item \textbf{Quantum immune:}
The reliance of blockchain technology on classical digital signatures makes it prone to quantum attack due to which an attacker equipped with a quantum computer can forge a digital signature or impersonate that user~\cite{fedorov2018quantum, suhail2020role}.
Realizing the fact arises due to quantum nervousness, IOTA uses the Winternitz One-Time Signature Scheme (WOTS)~\cite{buchmann2011security} which is a quantum-resistant algorithm. Nonetheless, there is an ongoing debate on the ability of blockchain to survive against quantum computers~\cite{QRBC2018, QRBCJun2018, casino2018systematic} and efforts are underway for quantum-safe blockchains~\cite{fedorov2018quantum}. 
\end{enumerate}
\begin{table}[htb!]
\begin{center}

 \begin{tabular}{c c } 
  \hline
\textbf{Symbol} & \textbf{Description} \\ [0.5ex] 
 \hline
 \hline
 SC & Supply Chain  \\
 ESC & Electronic Supply Chain  \\
 QC & Quality Control  \\
 $T_{ID}$, $B_{ID}$, $C_{ID}$, $S_{ID}$ & Transaction ID, Batch ID, Component ID, Sensor ID \\
 $Payload_a$, $Payload_f$ & Attach payload, Fetch payload\\
 $D_p$ & Data publisher or Seller  \\ 
 $D_r$ & Data receiver or Buyer  \\ 
 $\mathcal{K}$ & Authorization key \\
  $\mathcal{K}_{pu}$,  $\mathcal{K}_{pr}$ & Public key pair, Private key pair \\
 $T_{Data}$, $A_{Data}$ & Transaction data, Auxiliary data  \\ 
 $P_{Data}$, $P_{collect}$, $P_{aggr}$ & Provenance data, Provenance collect, Provenance aggregate \\ 
 $T_{Data}$, $A_{Data}$, $S_{Data}$ & Transaction data, Auxiliary data, Sensor data \\
 $Con_{info}$ & Consignment information \\
 $Reg_{cert}$ & Certificate by regulatory authority \\
 $s\_{d}$ & data from sensor devices \\
 $Src_{ID}$, $Prev_{TID}$ & Source ID, Previous Transaction ID \\
\hline
\end{tabular}
\caption{List of notations}
\label{tab:symbols}
\end{center}
\end{table}

\section{System Model} \label{SystemModel}
In this section, we provide an overview of the design parameters necessary for the SC system. We describe the network model and the data model that we consider for our proposed IOTA-based provenance scheme for SC. We also present the provenance model along with the outline of elemental provenance data components that are utilized in our proposed scheme. Finally, we discuss the security goals that our proposed scheme aims to achieve. Table~\ref{tab:symbols} lists the major symbols used throughout the paper. 
\begin{figure}[!hbt]
    \centering
\begin{tikzpicture}[
   planet/.style = {circle, draw=magenta, semithick, fill=magenta!10,
                    font=\large\bfseries, 
                    text width=22mm, inner sep=1mm,align=center}, 
satellite/.style = {circle, draw=#1, semithick, fill=#1!30,
                    text width=18mm, inner sep=1mm, align=center},
      arr/.style = {-{Triangle[length=3mm,width=6mm]}, color=#1,
                    line width=3mm, shorten <=1mm, shorten >=1mm}
                    ]
\node (p)   [planet]    {Challenges in SC};
\foreach \i/\j [count=\k] in {magenta/Integrity, magenta/Visibility, magenta/Scalability, magenta/Confident-iality, magenta/Data accessibility, magenta/IoT\&IIoT, magenta/Traceability, magenta/Risk, magenta/Data validation, magenta/Legacy systems,magenta/Disparate sources, magenta/Data availability, magenta/Automation}
{
    \node (s\k) [satellite=\i] at (\k*30:3.6) {\j};
    \draw[arr=\i] (p) -- (s\k);
}
    \end{tikzpicture}
\caption{Technical and non-technical challenges in the supply chain.} \label{fig:features}
\end{figure}
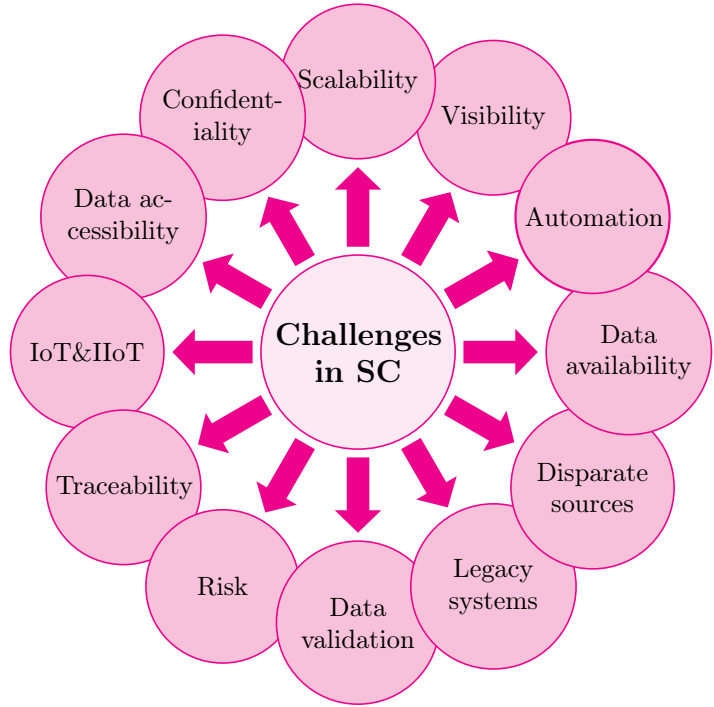

\subsection{Design Approach}
In this subsection, we highlight the technical and non-technical factors that must be considered before designing a provenance-based solution for SC.
We also provide a succinct overview of how these design principles can solve potential SC challenges (shown in Fig. \ref{fig:features}). 

The \textit{first} factor is to identify the information type and source, for instance, level of information (coarse-grained or fine-grained), data acquiring source such as digital assets or humans (each having different repercussions), etc. Comprehensive data aggregated from multiple data sources plays a significant role in solving many problems such as data traceability, risk factors in trade events, etc.

The \textit{second} factor is to identify erroneous data in the system to ensure data trustworthiness which in turn can solve many issues such as bullwhip effect, GIGO problem, trust issues between seller-buyer pair, etc. Erroneous data refers to the state of data before it arrives at the ledger, i.e., during data generation and data transit. Erroneous data can be generated (either maliciously or mistakenly) by (i) source/data originator, (ii) intermediate entities, (iii) SC participating entity, and (iv) sensors or other technologies connecting the physical and digital world. 

The \textit{third} factor is to identify the best practices for Supply Chain Risk Management (SCRM). SCRM involves processes to identify risk events and to activate a plan accordingly to mitigate its effect. Problems such as shrinkage, outage, natural disasters, economic crisis, etc. are covered under this factor.   

The \textit{forth} factor is the integration of state-of-the-art technologies such as IoT, IIoT, and others. Such integration can automate the processes of inventory with minimal human involvement. 

The \textit{fifth} factor is the evaluation of non-technical factors such as Ethical, Sustainable, and Responsible (ESR) operations as discussed in \cite{babich2019blockchain}. ESR operations deal with issues, such as labor conditions, child labor, responsible usage of natural resources (land, water, energy), etc. 

Note that, the above-mentioned factors are tightly inter-linked with each other, i.e., failure to exercise any one of the factors can highly affect the outcome of the other factor.

\begin{figure}[!ht]
\centerline{\includegraphics[width=4.0in]{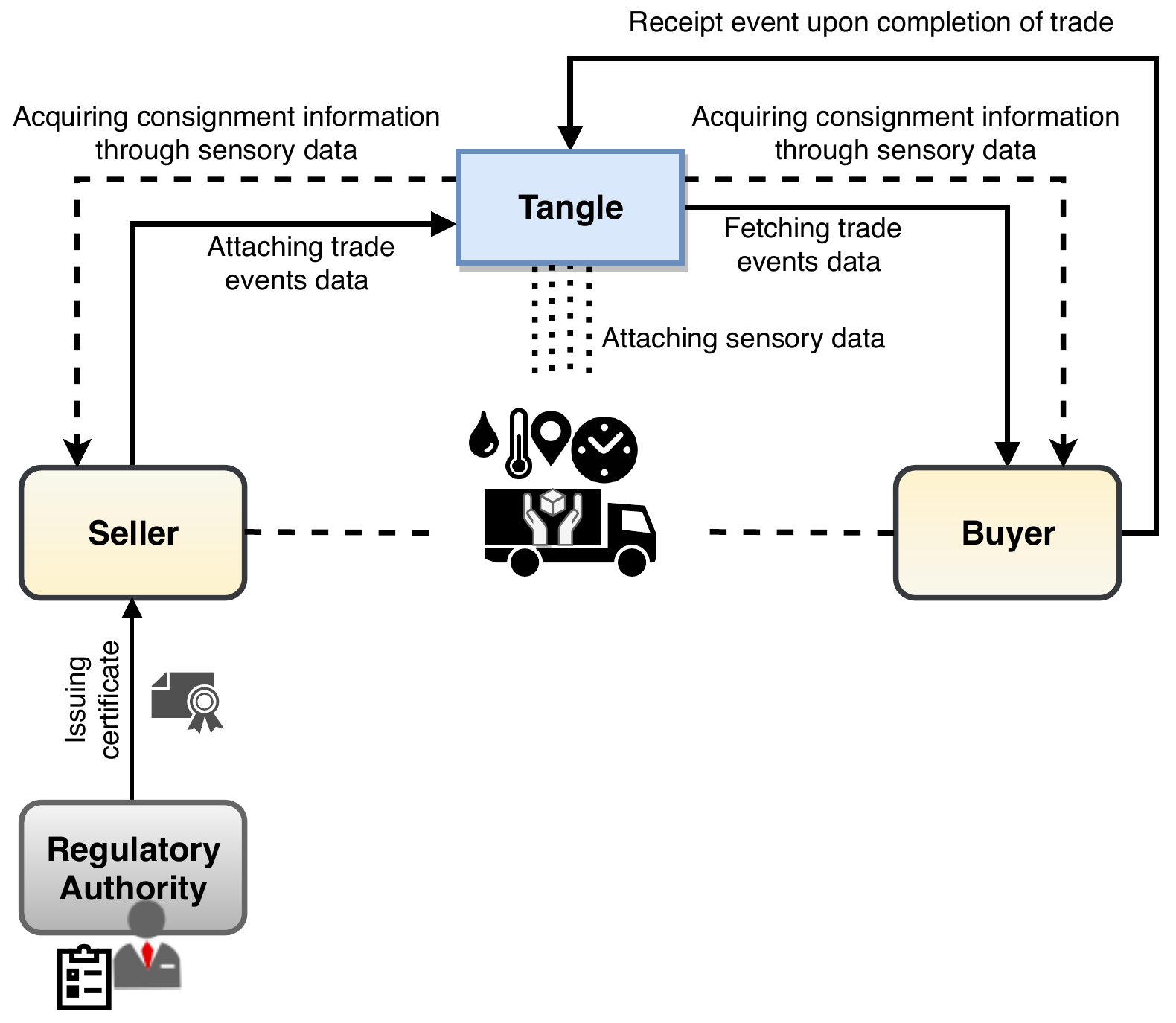}}
\caption{Network model illustrating a trade event between seller-buyer pair.} 
\label{fig:nwModel}
\end{figure}

\subsection{Network Model}
 Our network model consists of SC entities (participating and non-participating) and sensors. 
In the following, we outline two main data sources in the ESC, i.e., SC players and sensors.

\subsubsection{SC Players}
Following are the participating players in ESC. 

(i) Raw producers, (ii) Suppliers, (iii) Manufacturers, (iv) Warehouses, (v) Logistics, and (vi) Retailers.
The raw producers provide raw materials to the supplier to produce chip-sets and other peripherals. Those components are fabricated and assembled at a manufacturing unit. The finished products (for instance, mobile phones) are delivered to the warehouses for distribution. Finally, customers can purchase them from designated retailers. 

Additionally, there are non-participating players such as (i) customers, and (ii) researchers/analysts.
Non-participating members including customers and researchers are not involved in the SC process; however, they may need to fetch the production and manufacturing information about the products. Hence, they are also considered as part of our network model.

\subsubsection{Sensors}
Sensors are used to connect the physical world to the digital world. Sensors are affixed to batches during logistics and transportation to provide information such as location, temperature, humidity, etc. 
Due to the resource-constrained nature of sensors, we assume that such devices act as light nodes and may utilize the full nodes for performing computationally expensive tasks of the IOTA framework.
For further processing, interpretation, and analysis of data, the sensor data is fetched from the tangle to track and trace SC events.

\begin{figure}[!ht]
\centerline{\includegraphics[width=2.5in]{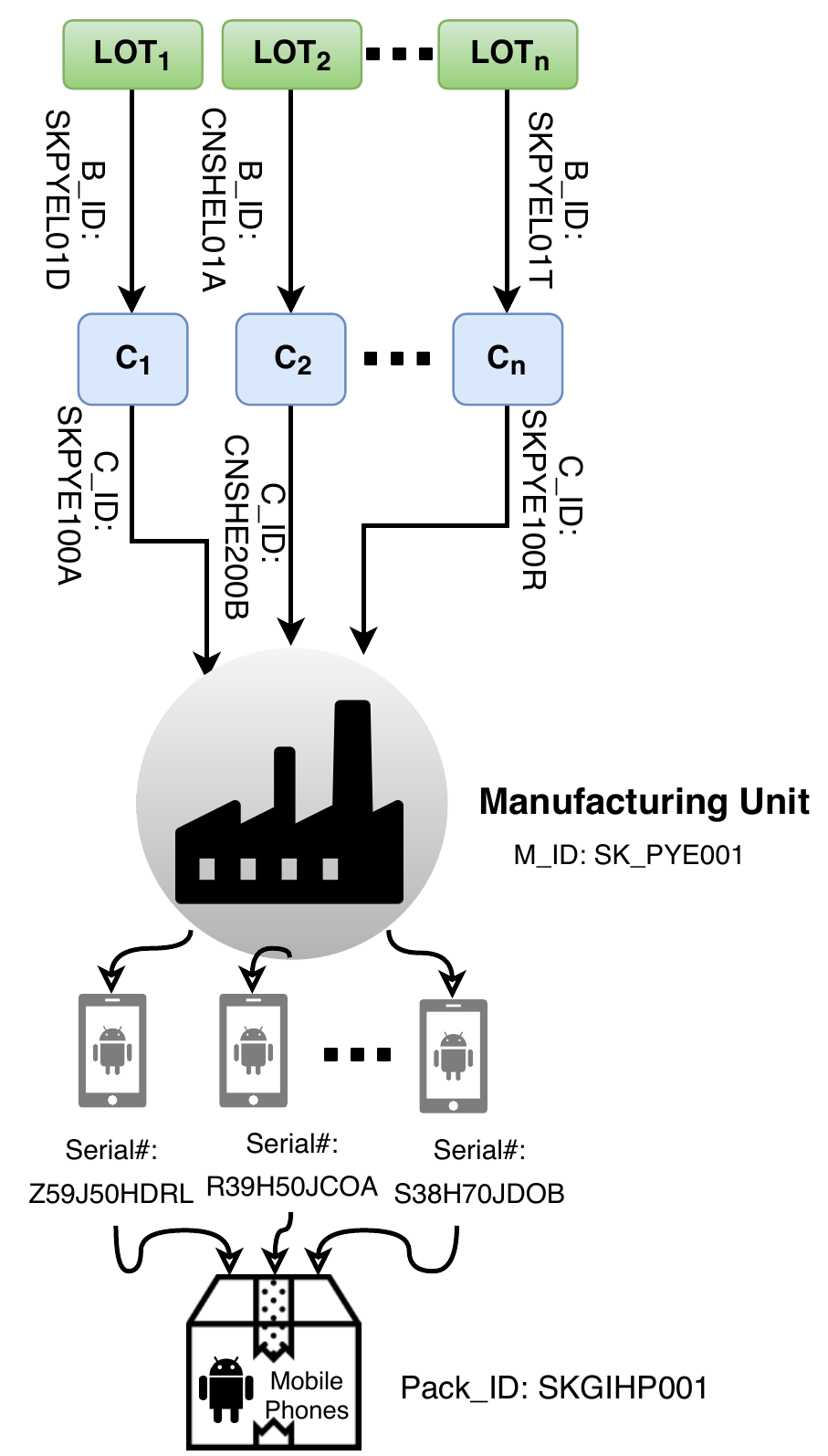}}
\caption{Product manufacturing and assembly at the manufacturing unit.}
\label{fig:M_unit}
\end{figure}

\subsection{Data Model} \label{data_model}
We assume that each SC entity acts as a \textit{Data Publisher} ($D_p$) and publishes its data (also referred to as \textit{attaching data}) on its \textit{MAM channel} identified by $Channel_{ID}$. On the other hand, the interested viewers act as a \textit{Data Receiver} ($D_r$) and subscribe to the desired channel ($\mathcal{C}$) to gain access to the data (also referred to as \textit{fetching data}) by using an authorization key ($\mathcal{K}$). The term $\mathcal{K}$ is collectively used for public $\mathcal{K}_{pu}$ and private $\mathcal{K}_{pr}$ key pairs. 

In the context of SC, $D_p$ and $D_r$ can be referred to as \textit{Seller} and \textit{Buyer} respectively.
The data (also referred as payload) consists of (i) Transaction data ($T_{Data}$), (ii) Auxiliary data ($A_{Data}$), and (iii) Sensor data ($S_{Data}$) can be represented as:

\begin{subequations}\label{1}
\begin{equation}\label{1a}
Payload \leftarrow T_{Data}||A_{Data}||S_{Data},
\end{equation}
\begin{equation}\label{1b}
T_{Data} \leftarrow T_{ID} || Seller_{ID} || Buyer_{ID} || Con_{info},
\end{equation}
\begin{equation}\label{1c}
A_{Data} \leftarrow QC || Reg_{cert}|| optional\_field,
\end{equation}
\begin{equation}\label{1d}
S_{Data} \leftarrow S_{ID} || Channel_{ID} || s\_d || timestamp.
\end{equation}
\end{subequations}

$T_{Data}$ consists of transaction ID ($T_{ID}$), trade event as $\langle source, destination \rangle$ pair, i.e., seller ID ($Supplier_{ID}$) and buyer ID ($Buyer_{ID}$), and consignment information ($Con_{info}$).
$Con_{info}$ may include batch ID ($B_{ID}$), component ID ($C_{ID}$), make and model number (as depicted in Fig. \ref{fig:M_unit} and Fig. \ref{fig:productledger}). Other granular details such as quantity, unit price, vehicle ID, etc., can also be included as a part of $Con_{info}$.

$A_{Data}$ consists of Quality Control ($QC$) parameters (such as ISO certifications/accreditation, warranty, etc.), regulatory endorsements certificates ($Reg_{cert}$), and optional field ($optional\_field$).
We consider $optional\_field$ to store application-specific or user-specific information, for example, pre-defined agreements among trading entities. In our case, we use this field to store packaging information (i.e., traceability information) related to items that may need to be packed together and extracted at a later stage such as during assembling mobile phone parts or during burning software on ICs. Traceability information includes \textit{where} a particular package is reopened by \textit{which} entity due to \textit{what} reason. Besides, it may contain information about shrinkage events in case of loss or damage to physical goods. 
Among other quality control parameters, warranty plays an effective role as it provides a marketing strategy to attract customers and also signals product quality~\cite{chen2012manufacturer}. In our case, this factor can also contribute to establishing a trust relationship among buyers (consumers) and sellers (honest or dishonest) in the long run.
The regulatory endorsements can help to ascertain that SC processes are abiding by ethical practices and environmentally-responsible operations. The exercising of such ESR operations requires the involvement of regulatory bodies (NGOs, governments, industry self-regulators) to conduct a periodic on-site inspection of the units and provide verifiable certificates ($Reg_{cert}$) as shown in Fig. \ref{fig:nwModel}.

$S_{Data}$ consists of sensor ID ($S_{ID}$) assigned to each sensor, $Channel_{ID}$ of $D_p$ publishing the sensor data ($s\_d$) such as location, temperature, humidity along with timestamp information. During transportation and logistics of goods, $s\_d$ is attached to the tangle and can be accessed by seller-buyer pair to acquire $Con_{info}$ as shown in Fig. \ref{fig:nwModel}.
The granularity level of $s\_d$ can be customized depending on the requirements, for instance, coarse-grained data by averaging temperature data or fine-grained data by using channel splitting option. 

When a product or its parts are received by the buying entity, the receipt ($Receipt$) is generated to log the completion of the trade event between the seller-buyer pair (as shown in Fig. \ref{fig:nwModel}).
\begin{equation}\label{eq:receipt}
Receipt \leftarrow T_{ID} || status,
\end{equation}
where $T_{ID}$ and $status$ represent the transaction ID and status of the received item respectively. The purpose of introducing this transaction is threefold: (1) to keep track of the successful transactions to avoid any fake-progressive sub-chains, (2) to indicate any loss or damage event, and (3) to integrate trade finance process in SC.

\subsection{Provenance Model}
Deriving the product story primarily involves collecting provenance data ($P_{Data}$) based on $\langle source, destination \rangle$ pairs while traversing through the SC process. Therefore, to construct and assemble $P_{Data}$, firstly the payload (holding complete transaction and auxiliary data) is fetched from the ledger and secondly the required information is acquired from the fetched payload ($Payload_f$). 
The key factors to devise product provenance are:
\begin{equation}\label{eq:pro}
P_{Data} \leftarrow Channel_{ID}|| T_{ID} || Src_{ID} || Prev_{TID},
\end{equation}
where $Channel_{ID}$ refers to the current ID of source, $Src_{ID}$ refers to the channel ID of destination (i.e., immediate $Channel_{ID}$ of SC entity), and $Prev_{TID}$ refers to the on-going transaction in the channel of $Src_{ID}$ pertaining to the fact that there can be multiple on-going transactions in that channel. Note that depending on the query, additional information can be obtained from $T_{Data}$, $A_{Data}$, and $S_{Data}$ accordingly. Also, $P_{Data}$ can be encoded on a QR code to be used by SC entities.

\subsection {Security Goals} \label{security_goals}
Keeping in view the requirements of SC, our proposed scheme aims to achieve the following security properties:

 \begin{enumerate}
 \item \textbf{Data confidentiality:} 
To hide classified trade information among competitors, it is essential to encrypt data communication.
\item \textbf{Access control rights:}
 Defining access control rights is indispensable to conceal classified trade information among competitors. Moreover, sharing only a subset of data at any desired point in time must be allowable by the SC player.
\item \textbf{Restrictions on data retrieval:}
Upon joining the data stream of the SC process, SC players must only be able to retrieve the information at or after their entry point to the process with no privileges to previous transaction streams.

{{ \item \textbf{Data integrity:}
Integrity of trade events must also be ensured during data creation and sharing.  
}
{ \item \textbf{Non-repudiation:}
Any participating entity must not be able to deny an SC event that has happened or SC data that has been produced.}
}
\end{enumerate}

\section{Proposed Framework: Procuring Provenance in ESC Through IOTA} \label{scheme}
In this section, we provide a brief overview of the characteristics and working of the MAM protocol provided by IOTA. We also devise the proposed framework for provenance in SC using IOTA with the help of algorithms and flow diagrams. 

\subsection{Masked Authenticated Messaging (MAM)} \label{MAM}
To ensure secure, encrypted, and authenticated data stream on the tangle, we leverage a MAM module that provides a channel where data owner who publishes the data and data viewers who subscribe, meet. Using the gossip protocol, the message from the data publisher is propagated through the network and can be accessed by the channel subscribers only.

\subsubsection{Generating Message Chain}
\begin{figure}[!ht]
\centerline{\includegraphics[width=4.0in]{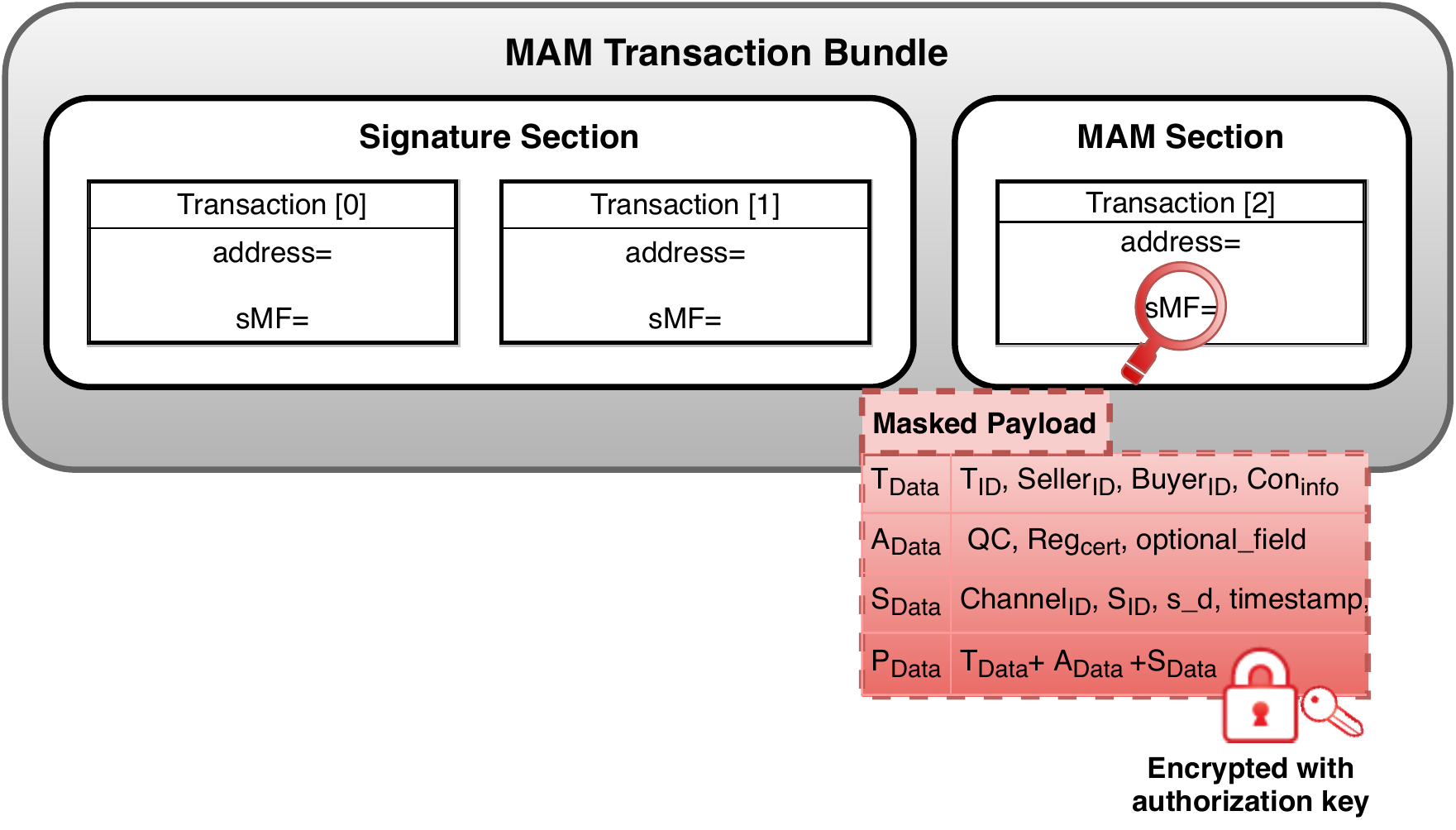}}
\caption{MAM transaction bundle illustrating \textit{signature section} and \textit{MAM section}. MAM section consists of \textit{Masked Payload} encrypted with authorization key (restricted channel mode). Masked Payload consists of transaction data $T_{Data}$, auxiliary data $A_{Data}$, and sensor data $S_{Data}$ whereas $P_{Data}$ can be constructed from $T_{Data}$, $A_{Data}$, and $S_{Data}$.}
\label{fig:mam_working}
\end{figure}

A MAM transaction bundle consists of two sections including (i) \textit{Signature}, (ii) \textit{MAM}. \\Fig. \ref{fig:mam_working} shows the main components of \textit{MAM Transaction Bundle}.

The \enquote{MAM section} contains the masked message. To post a masked message, MAM deploys \textit{Merkle tree-based signature scheme} \cite{Merkle1988} that requires the creation of a root to view the payload. Furthermore, to support forward transaction linking, the MAM section also contains a connecting pointer i.e.,\textit{nextRoot} and other associated entities that are required for fetching the next payload. The approach to access the payload depends on the channel mode used, for instance, restricted channel mode requires authorization key pairs to encode and decode messages. 

For the validity check of the MAM section, data publishers add a signature in the MAM bundle and store it in the \textit{signature Message Fragment(sMF)} of the transaction. Such transactions are stored in \enquote{Signature section} of the MAM bundle. A comprehensive working of the MAM protocol is explained in~~\cite{coinmonks}. 

\subsubsection {Access Control and Provision of Authenticated Data}
To control the data accessibility and visibility in the tangle, MAM provides the following channel modes: (i) \textit{public}: address = root i.e., by using the address of the message, any random user can decode it, (ii) \textit{private}: address = hash(root) i.e., the hash of the Merkle root is used as the address, thus, preventing random users from deciphering message as they are unable to derive the root from the hash, and (iii) \textit{restricted}: address = {\it hash(root) + authorization key} i.e., the hash of the authorization key and the Merkle root is used as address, thereby allowing only authorized parties to read and reconstruct the data stream. Changing the authorization key results in revoking permission to access the data without requiring the data publisher to change its $Channel_{ID}$. It is important to note that considering the confidential trade flow requirements of the SC players, we prefer the use of restricted channel mode of MAM. Furthermore, to enforce the ownership of the channel, signature validation is performed upon message reception to authenticate the source of the message or in other words to validate the ownership of the publisher. Failure to signature verification results in an invalid message.

\subsection{Provenance in ESC}
In this subsection, we provide a detailed description of the proposed provenance-based SC framework.

\subsubsection{Seed Generation}
To initiate the communication process based on the address and private key requires a seed. A seed is more like a private key and consists of 81 characters including upper case alphabets and digit 9. The seed generation process uses environmental noise (for example, device drivers, network packet timing, etc.) as an input to a Cryptographic Secure Pseudorandom Number Generator (CSPRNG), to produce random seed values. 

\subsubsection{Setting Security Level, Channel Mode, and Key Generation}
IOTA has defined three security levels: 1 (low), 2 (medium), or 3 (high). The default security level is 2; however, we use the recommended security level 3. To keep the communication confidential, we set the channel mode as ``\textit{restricted}'' so that the authorized parties can access the data based on shared authorization key pairs. $\mathcal{K}$ is used to encrypt and decrypt the payload by a sender and receiver entities, respectively. The cryptographic keys can be shared among the participating parties by using any of the existing key exchange technique, for instance, Elliptic Curve Cryptosystems (ECDSA~\cite{johnson2001elliptic} or ECDH~\cite{ECDH}). However, the existing key exchange systems have an Achilles’ heel due to the not-so-far quantum computing attacks, therefore, a lattice-based public-key cryptosystem N{\it{th}} Degree Truncated Polynomial Ring (NTRU)~\cite{ntru} must be adopted as it allows to generate and exchange key pairs in a quantum secure way. In our case, we use the NTRU key exchange protocol.

\subsubsection{Data Publishing}
Each SC player creates a channel $\mathcal{C}$ to publish its data on the tangle. For further details of the payload and its sub-entities, we refer to Section \ref{data_model}. Upon selection of channel mode, security level, and authorization key, finally, the MAM transaction bundle is attached ($Payload_a$) to the tangle. Algorithm \ref{algo:data_pub} illustrates the steps of data publishing.

\begin{algorithm}[!ht]
\caption{Data publishing}\label{algo:data_pub}
\begin{algorithmic}[1]
\Require $seed, root$
\Ensure $Payload_a$
\State mamState $\leftarrow$ Mam.init (iotaObject, seed, securityLevel) 
\State mamState $\leftarrow$ Mam.changeMode (mamState, channelMode, $\mathcal{K}_{pub}$) \Comment{Set channelMode as `restricted' and use public key pair to encrypt the payload.}
\State MAMObject $\leftarrow$ Mam.create (mamState, payload) \Comment{Create MAM payload which consists of transaction and auxiliary data.}
\State Mam.attach (MAMObject.payload, MAMObject.address) \Comment{Attach the payload to the tangle.}
\end{algorithmic}
\end{algorithm}

\subsubsection{Data Receiving}
The interested SC players subscribe to the channel to view the published data. The subscribers are able to receive or fetch payload ($Payload_f$) based on root and decipher the payload based on $\mathcal{K}$ as presented in Algorithm \ref{algo:data_rec}.

\begin{algorithm}[!ht]
\caption{Data receiving}\label{algo:data_rec}
\begin{algorithmic}[1]
\Require $root$
\Ensure $Payload_{f}$
\State mamState $\leftarrow$ Mam.init (iotaObject, seed, securityLevel) \Comment{Set seed value and securityLevel as used in Algo. \ref{algo:data_pub}.}
\State mamState $\leftarrow$ Mam.changeMode (mamState, channelMode, $\mathcal{K}_{pr}$) \Comment{Set channelMode and private key pair to decrypt the payload.}
\State Mam.fetch (root, restricted, $\mathcal{K}$) \Comment{Fetch message stream from the tangle.}
\end{algorithmic}
\end{algorithm}

Recalling the notion \enquote{IOTA for IoT}, it is important to mention that IOTA enables flexible integration of the sensor data ($S_{Data}$) in the tangle. Hence, $S_{Data}$ can be published, fetched, and analyzed following a similar approach as that of data publishing and data receiving. For instance, the consignment information can be acquired by the seller or buyer as shown in Fig. \ref{fig:nwModel}.

 \begin{figure}[!hb]
\centerline{\includegraphics[width=5.8in]{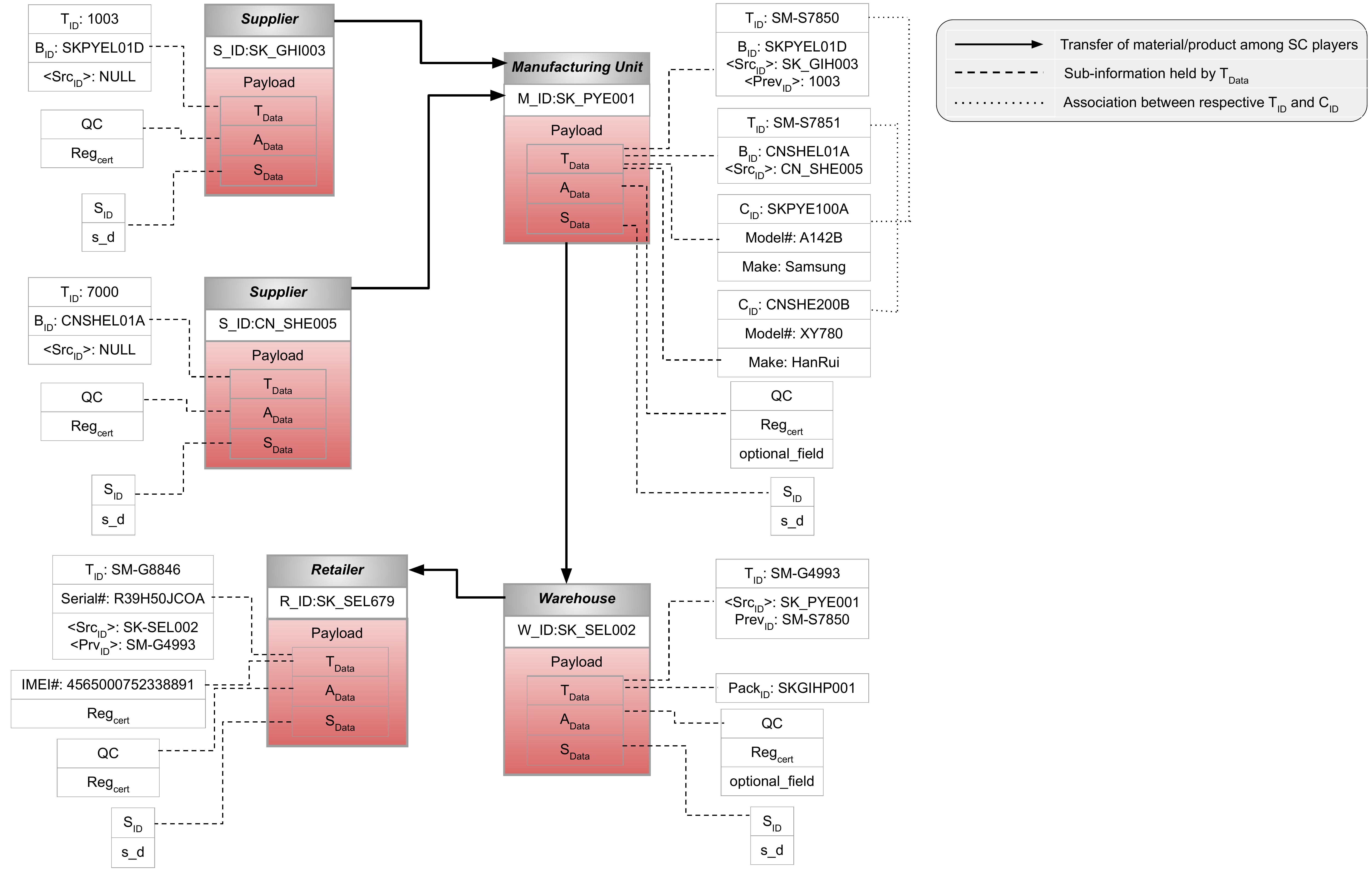}}
\caption{Product ledger: an example scenario illustrating the process of collection and aggregation of \textit{provenance data} throughout the ESC.} 
\label{fig:productledger}
\end{figure}

\begin{algorithm}[!ht]
\caption{Collecting and aggregating provenance data from the fetched payload}\label{algo:pro_data}
\begin{algorithmic}[1]
\Require $root$
\Ensure $P_{Data}$
\Procedure{fetch\_aggr($P_{Data}$)} {}
\Do
\For{\texttt{each subscribed channel $\mathcal{C}_{i}$}} 
\State Mam.fetchSingle (root, restricted, $\mathcal{K}$) \Comment{Fetch transaction from the subscribed channel.}
       \State $Payload_{f}$= $T_{Data} || A_{Data} || S_{Data}$ \Comment{Fetch and decipher the payload.} 
      \State $P_{collect}$ $\leftarrow$ $Channel_{ID}$ || $T_{ID}$ || $Src_{ID}$ || $Prev_{TID}$   \Comment{Collect provenance data from fetched\_payload.} 
      \State $P_{aggr}$ $\leftarrow$  $P_{collect}$ || $A_{Data}$ || $S_{Data}$ \Comment{Aggregate other granular details (if required).} 
      \State $P_{Data}$ $\leftarrow$ $P_{aggr}$
      \State goto $\langle Src_{ID} \rangle$ channel \Comment{Go to the intermediate source channel.}
  \State look for $Prev_{TID}$ == $T_{ID}$  \Comment{Look up for the transaction ID.}

\EndFor
 \doWhile{$\langle Src_{ID} \rangle$  $\neq$ NULL} 
\EndProcedure
\end{algorithmic}
\end{algorithm}

\subsubsection{Collecting and Aggregating Provenance Data}
Collecting $P_{Data}$ consists of three steps: (i) fetching payload ($Payload_f$), (ii) collecting provenance ($P_{collect}$), and (iii) aggregating provenance ($P_{aggr}$). Firstly, the payload is fetched from the tangle. Secondly, upon fetching the payload, $P_{collect}$ collects the information using key identifiers as mentioned in eq.~\ref{eq:pro}. Thirdly, the collected provenance information is maintained as $P_{aggr}$ along with other granular details (including consignment information, timestamped sensor data, quality control information, etc.). Finally, $P_{aggr}$ is then stored as $P_{Data}$. 
$\langle Src_{ID} \rangle$ refers to the channel address of the SC player who publishes the data through transaction $Prev_{TID}$. Throughout the chain, $\langle Src_{ID} \rangle$  helps in locating back to the intermediaries and ultimately the originator. Hence, moving to the next channel to collect, and aggregate provenance information is based on $\langle Src_{ID} \rangle$ and $Prev_{TID}$ to obtain the respective transaction. The process of fetching and aggregating provenance continues until the supplier is found. The steps for collecting and aggregating provenance data from the payload are illustrated in Algorithm \ref{algo:pro_data}. 

In order to explain the fetching and aggregating of $P_{Data}$, let us consider Fig.~\ref{fig:productledger}. Suppose that a SC player (customer or analyst) wants to trace back the product journey. Firstly, key identifiers are fetched i.e., $T_{ID}$=SM-G8846, $Product_{ID}$=R39H50JCOA, and $Channel_{ID}$=R\_ID: SK\_SEL679 from the \textit{Retailer} channel. Secondly, fetched data and auxiliary data are aggregated and collected in $P_{Data}$. Thirdly, based on the $\langle Src_{ID} \rangle$=SK\_SEL002 and $Prev_{TID}$=SM-G4993, the provenance information (for instance, $Pack_{ID}$= SKGIHP001), is then fetched from next \textit{Warehouse} channel such that $Prev_{TID}$ equals $T_{ID}$. 
Similarly, following $Prev_{TID}$=SM-S7850 and $\langle Src_{ID} \rangle$=M\_ID: SK\_PYE001 the information related to batch $B_{ID}$, model\# and make is obtained from \textit{Manufacturing Unit} channel. Here we can see that the batches holding components may arrive from different suppliers located in different countries. Hence, based on $T_{ID}$=SM-S7850, $B_{ID}$=SKPYEL01D and $C_{ID}$=SKPYE100A information are obtained. Also $\langle Src_{ID} \rangle$=S\_ID: SK\_GIH003 is used to reach the respective \textit{Supplier} channel. Since $\langle Src_{ID} \rangle$=NULL, therefore, no further $channel_{ID}$ is required to fetch more information. It is important to note that the additional information can be collected and aggregated from $T_{Data}$, $A_{Data}$, and $S_{Data}$ based on the user's query (Step 7). The query results also depend on access privileges defined by the SC players. Furthermore, the data can be fetched from any channel by any of the participating entities at any instant of time by using provenance key identifiers. 

\begin{table}[!ht]
\begin{center}

\begin{tabular}{clll}
 \hline
\textbf{Platform name} & \textbf{CPU} & \textbf{CPU core} & \textbf{Number of cores}\\ [0.5ex] 
\hline
\hline
  Desktop machine & Intel Core i5-3330 & -  & 4 (per socket) \\
  Raspberry Pi 3B~\cite{RPI3B} & BCM2837 &  Cortex-A53 & 4 \\
\hline
\end{tabular}
\caption{Specifications for hardware platforms.} \label{tab:platform}
\end{center}
\end{table}

\section{Performance Evaluation} \label{simulation}
In this section, we evaluate the proposed IOTA-based provenance scheme for SC. Overall, we consider 4 IOTA operations including (i) create, (ii) PoW, (iii) attach the payload to the tangle, and (iv) fetch payload from the tangle. Among these operations, we focus on attaching and fetching latency metrics. We also analyze the security of the proposed scheme with respect to SC requirements. 

\begin{figure}[!ht]
\centerline{\includegraphics[width=3.0in]{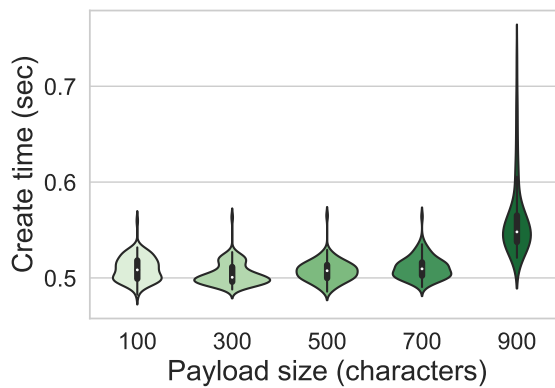}}
\caption{Estimated time (sec) required to create payload. }
\label{fig:create}
\end{figure}
\subsection{Simulation Setup}

\subsubsection{Hardware}
To evaluate the performance of the proposed scheme, we use a desktop machine and a Raspberry Pi 3B. The summary of hardware specifications for the test environment is shown in Table~\ref{tab:platform}.

\subsubsection{Software}
To evaluate IOTA operations, we use JavaScript and compile it for the considered target platforms. Both target platforms are running Linux operating system (Ubuntu 16.04 LTS). 
For operations including creating payload, attaching payload to the tangle, and fetching payload from the tangle, we use the current implementation of MAM protocol. We evaluate the proposed scheme for security level 3. 
For local PoW, we use CCurl\footnote{https://github.com/iotaledger/ccurl} library developed and maintained by the IOTA Foundation.
For local PoW, we use a PoW proxy server that acts as a dedicated proxy server to perform PoW for the targeted node. We also use a remote node selected from the available IOTA nodes list which is responsible for carrying out PoW on behalf of the targeted node. 

\subsection{Evaluation Metrics}
We put emphasis on the latency metric for the evaluation of IOTA operations. Each experiment is evaluated 100 times for security level 3. To represent the data distribution, we choose violin graph~\cite{hintze1998violin} that befits our representation requirements of the results. The violin graph indicates median (a white dot), quartiles (thick black bar) with whiskers reaching up to 1.5 times the inter-quartile range (thin black bar), and kernel probability density (colored area) that shows the distribution shape of data. With reference to the proposed scheme, parameters of the violin graph can be interpreted as follows: median represents the central value for performing IOTA operations (including creating, attaching, and fetching) whereas quartiles represent the overall range of data while performing IOTA operations.
Starting with payload creation, Fig.~\ref{fig:create} shows that the time required to create the payload is almost negligible. However, it is observed that for payload size 900 the distribution of data is different in comparison to others. The reason for such different behavior is particularly because of the creation of the bundle. In addition, if the payload size increases more than 2187 trytes (1300 characters) additional transactions in the bundle are required.

Next, we analyze the process of attaching payload to the tangle, fetching payload from the tangle, and PoW (local and remote) due to the fact that such IOTA operations have significant time delays in comparison to creating payload.   

\begin{figure}[!ht]
\centering
\begin{subfigure}[b]{0.48\textwidth}
\centering
\includegraphics[width=\textwidth]{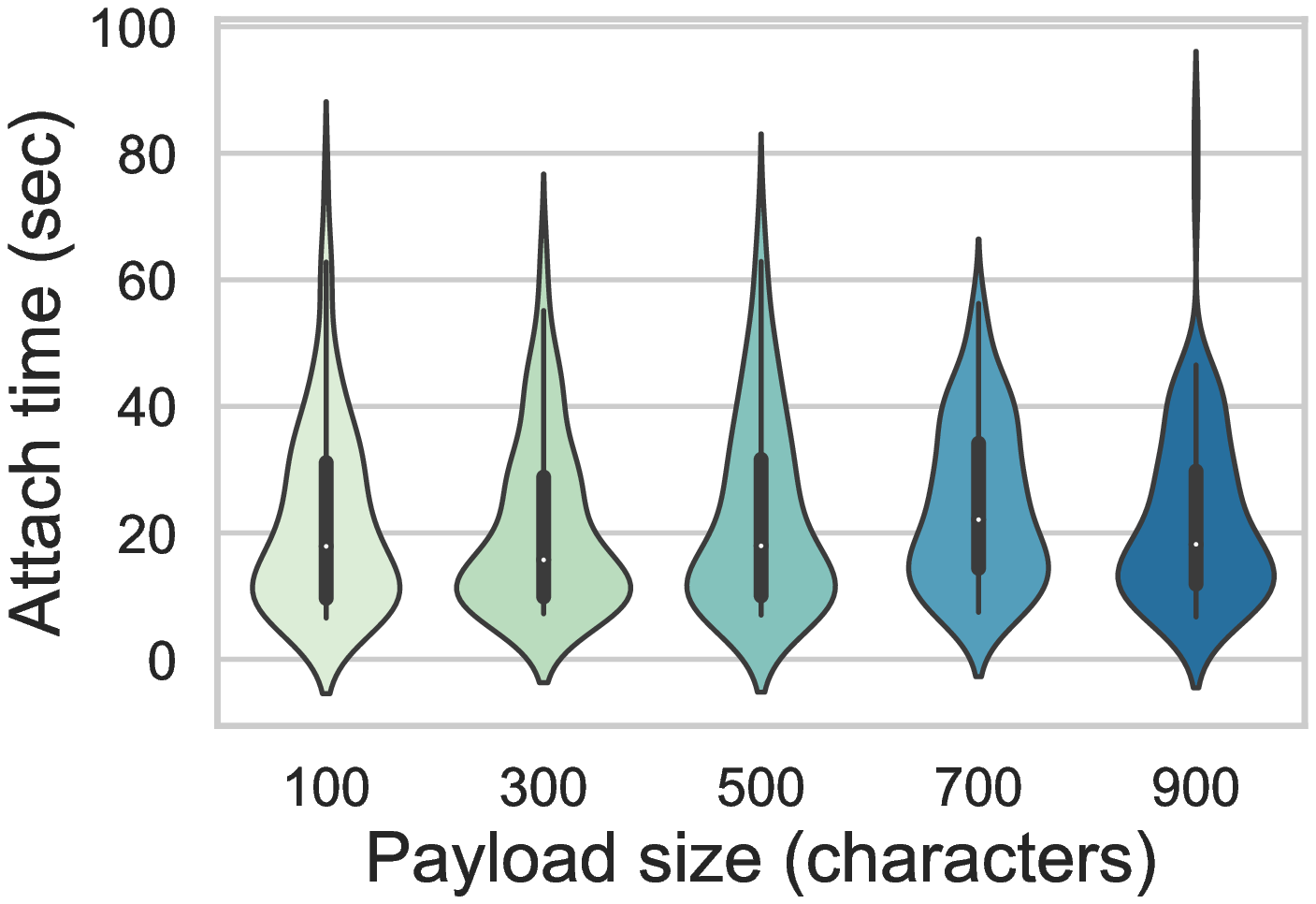}
\caption{}
\label{graph:attach}
\end{subfigure}%
\hspace*{\fill}
\begin{subfigure}[b]{0.48\textwidth}
\centering
\includegraphics[width=\textwidth]{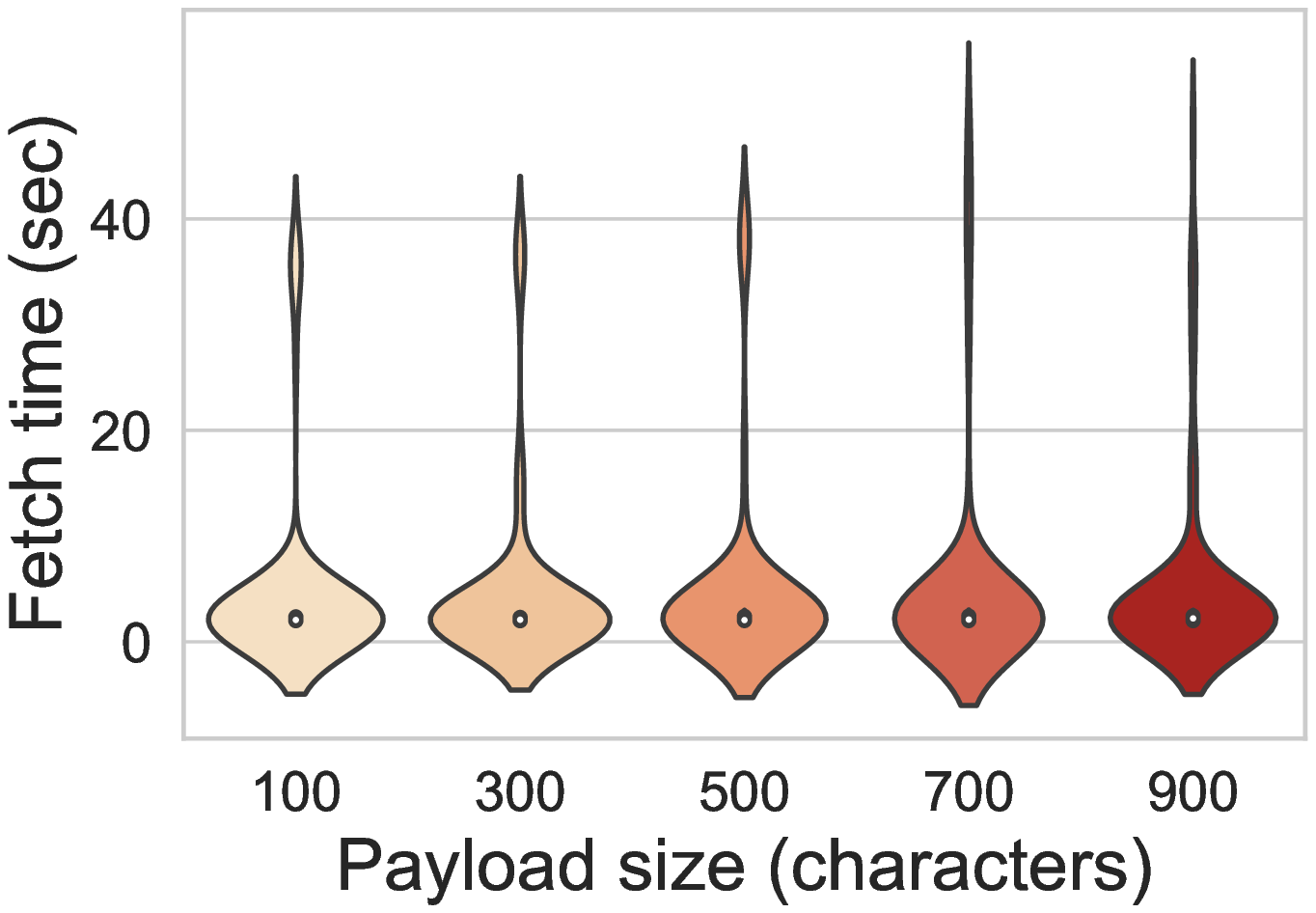}
\caption{}
\label{graph:fetch}
\end{subfigure}
\centering
\caption{Estimated time (in sec) required to (a) attach data to the tangle, (b) fetch data from the tangle by using remote node.}
\label{graph:remote}
\end{figure}

\begin{figure}[!ht]
\centering
\begin{subfigure}[b]{0.48\textwidth}
\centering
\includegraphics[width=\textwidth]{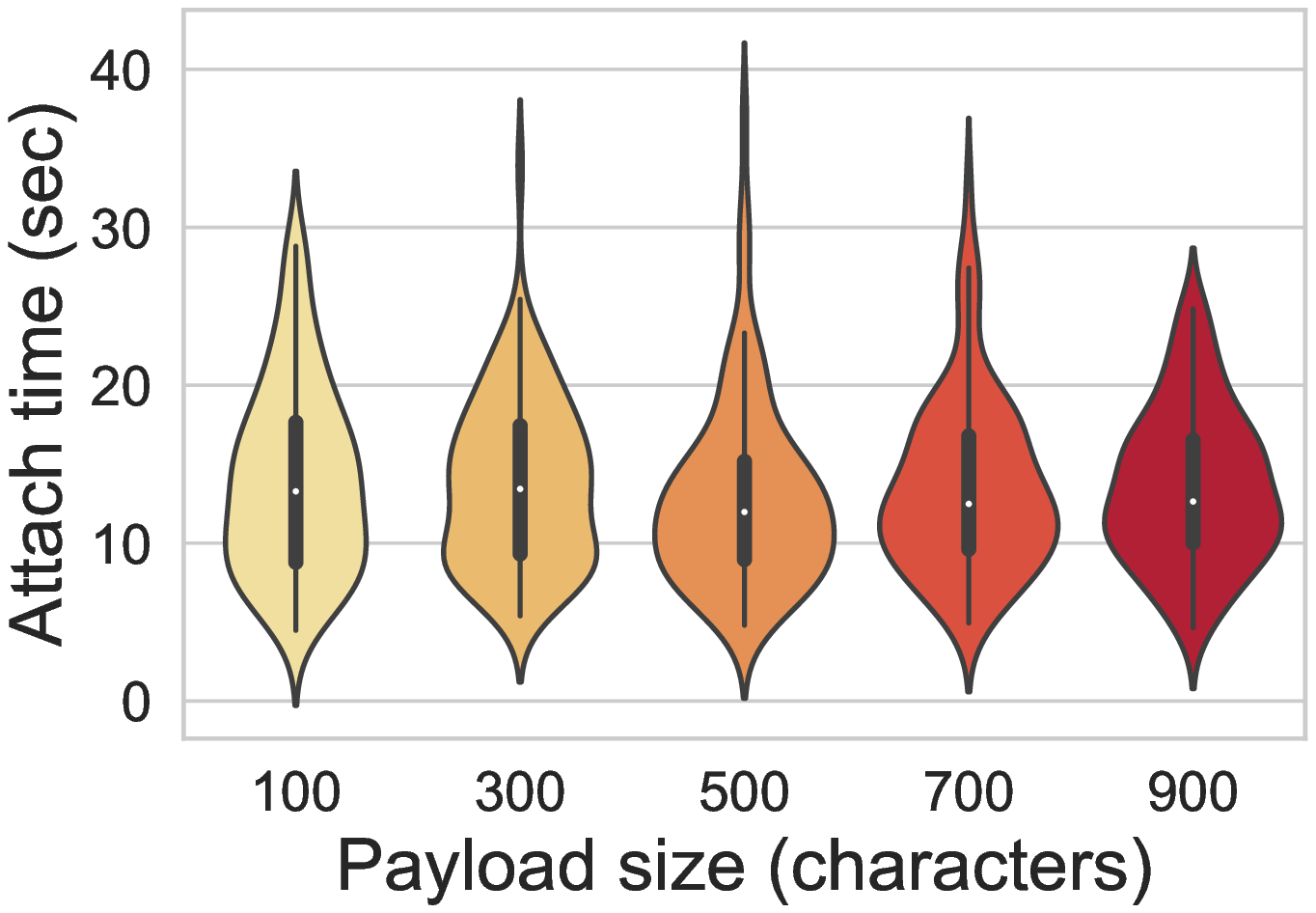}
\caption{}
\label{graph:attach_l}
\end{subfigure}%
\hspace*{\fill}
\begin{subfigure}[b]{0.48\textwidth}
\centering
\includegraphics[width=\textwidth]{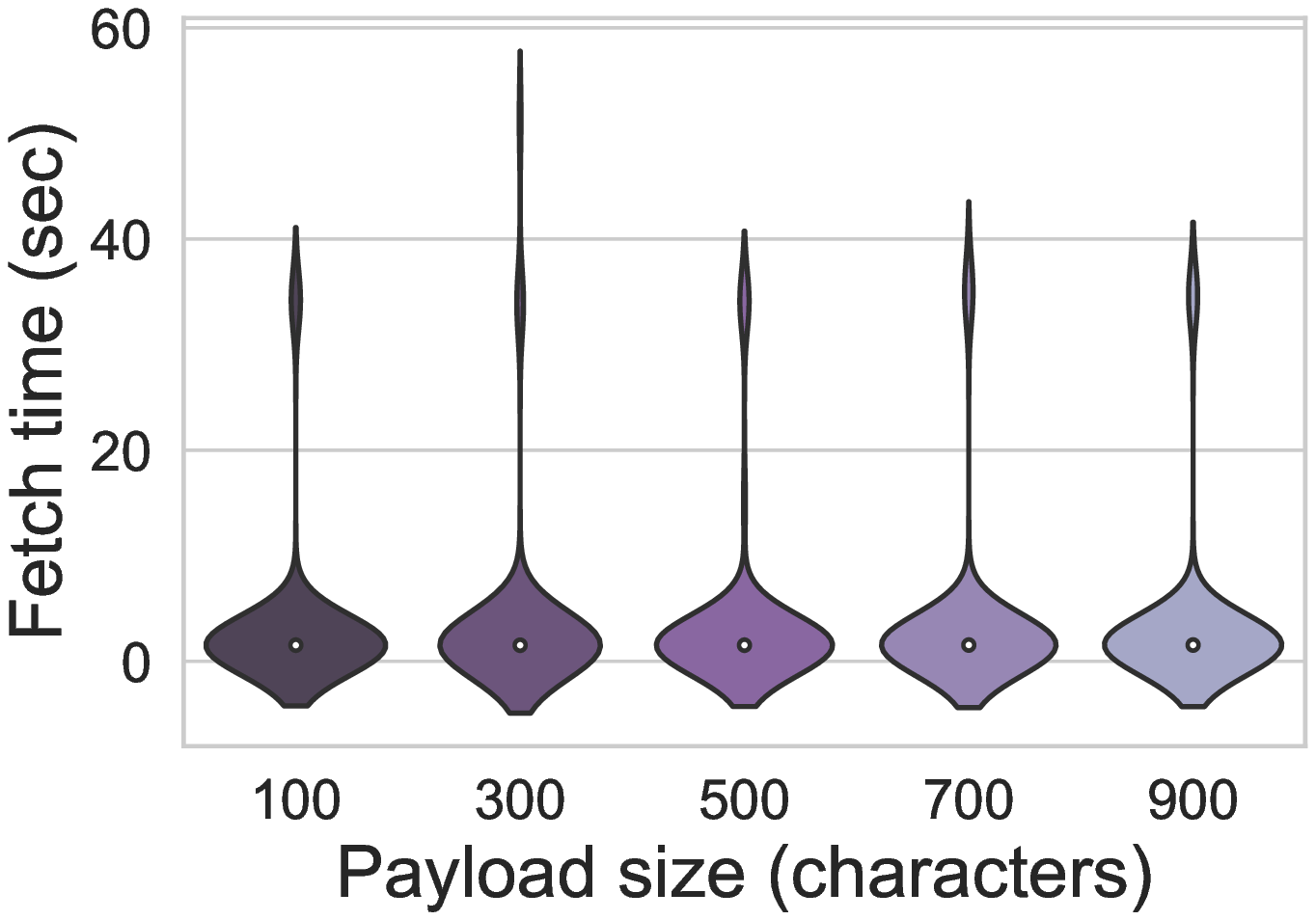}
\caption{}
\label{graph:fetch_l}
\end{subfigure}
\centering
\caption{ Estimated time (in sec) required to (a) attach data to the tangle, (b) fetch data from the tangle by using local node.}
\label{graph:local}
\end{figure}

It is important to note that the \textit{attach phase} corresponds to $Payload_a$ (attaching payload (consisting of $T_{Data}$ and $A_{Data}$)) while the \textit{fetch phase} corresponds to $Payload_f$ (fetching payload) from which $P_{Data}$ can be constructed. Depending on the query criteria and access privileges defined on the basis of channel splitting, $P_{Data}$ can be constructed. For simplicity, we consider that the query acquires every possible detail (i.e., entire payload) during the fetch phase based on which the provenance information can be derived upon the request of participating and non-participating entities. 

\begin{figure}[!ht]
\centering
\begin{subfigure}[b]{0.48\textwidth}
\centering
\includegraphics[width=\textwidth]{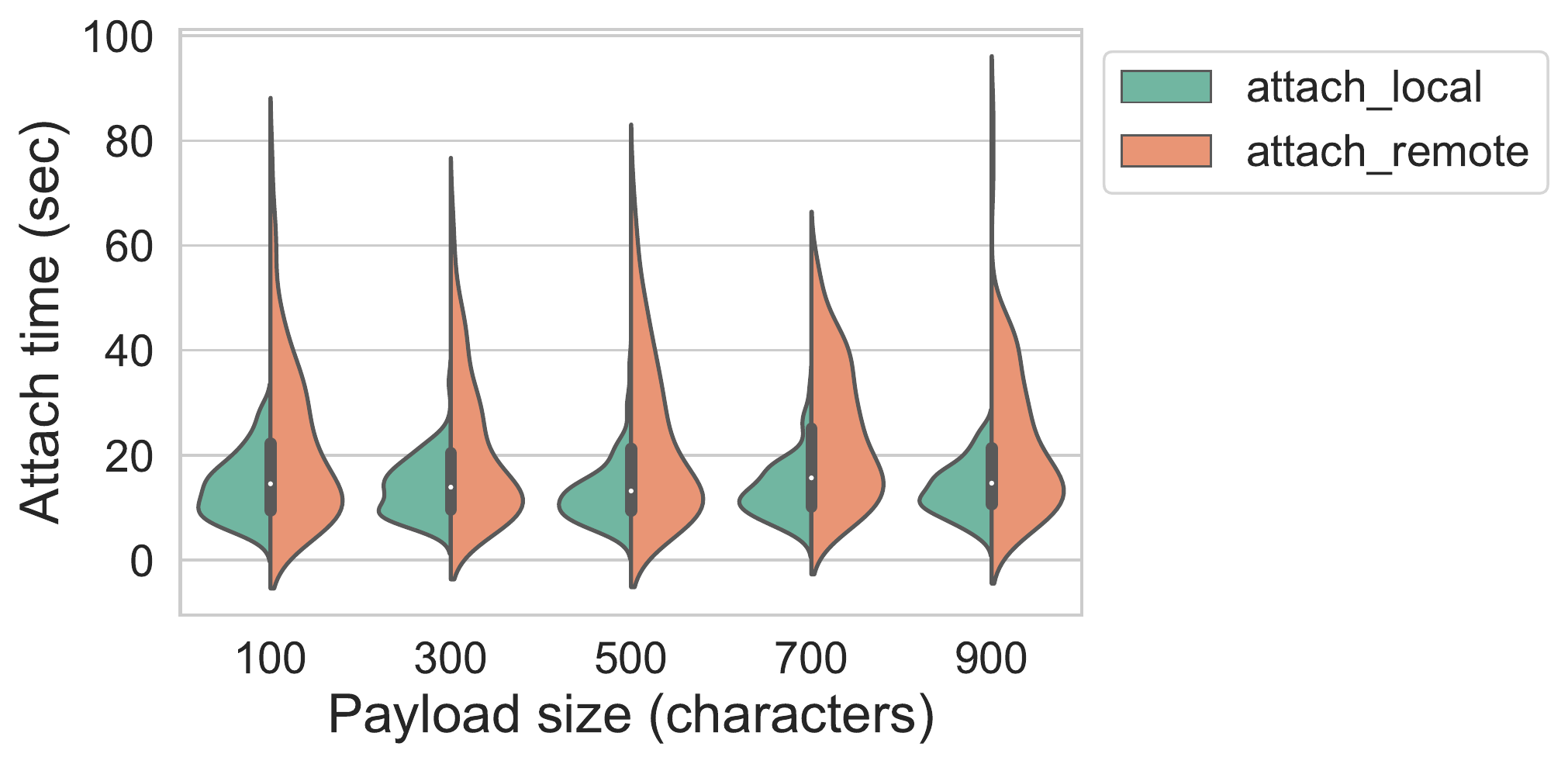}
\caption{}
\label{graph:split_attach}
\end{subfigure}%
\hspace*{\fill}
\begin{subfigure}[b]{0.48\textwidth}
\centering
\includegraphics[width=\textwidth]{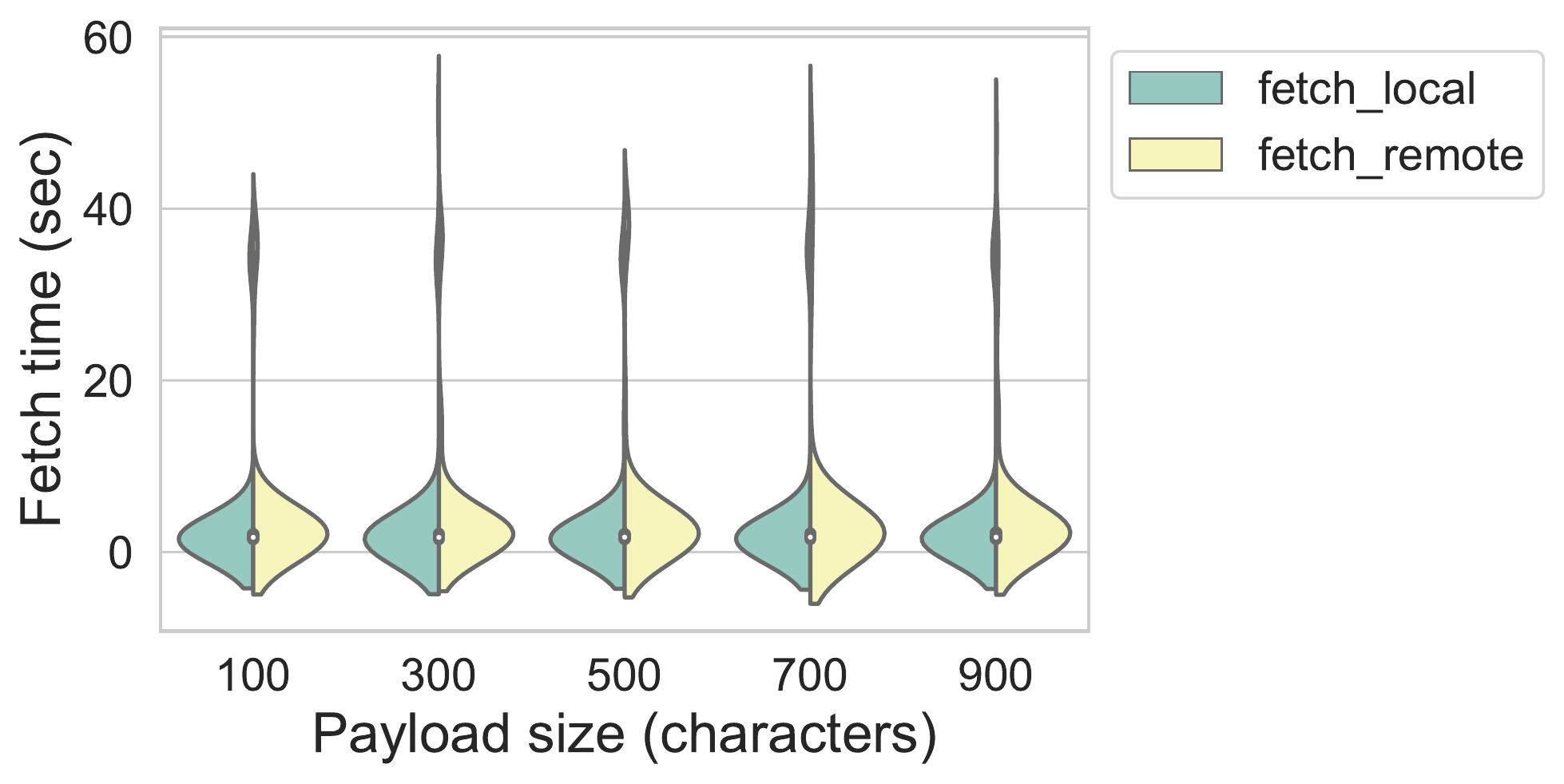}
\caption{}
\label{graph:split_fetch}
\end{subfigure}
\centering
\caption{Comparison between processing time required by local and remote nodes: estimated time (in sec) required to (a) attach data to the tangle, (b) fetch data from the tangle.}
\label{graph:split}
\end{figure}

Keeping in view the above-mentioned definitions of attaching and fetching data, firstly, we perform the attaching and fetching of payload (shown in Fig.~\ref{graph:remote}) by relying on a remote node and secondly, we perform the attaching and fetching of payload (shown in Fig.~\ref{graph:local}) by using a local node. In either of the cases, it is observed that the attaching process consumes more time in comparison to the fetching process. Both the attaching and fetching phases are independent of the payload size ranging from 100 to 900 characters. 
Similarly, we compare the attach and fetch process with respect to remote and local nodes. Figure~\ref{graph:split} shows that relying on a remote node to perform PoW incurs time delays and hence consumes time, whereas, local PoW does not incur much delays. 
In particular, Fig.~\ref{graph:split} represents the distribution of data for performing IOTA operations remotely and locally such that the median value for attaching payload is around 17 sec and 12 sec respectively, while the median value for fetching data is around 2 sec and 1 sec respectively.

\begin{table}[!ht]
\begin{center}
\begin{tabular}{cll}
 \hline
\textbf{Operation} & \textbf{Average Time (sec)} & \textbf{Energy Consumption (J)} \\ [0.5ex] 
 \hline
 \hline
Attach  & 23.1 & 5.1  \\
Fetch  & 6.4  &  1.4 \\ 
\hline
\end{tabular}
\caption{Average time measured and energy consumption estimated for attaching and fetching data from the tangle through Raspberry Pi 3B.} \label{tab:energy}
\end{center}
\end{table}

\begin{figure}[!ht]
\centering
\begin{subfigure}[b]{0.48\textwidth}
\centering
\includegraphics[width=\textwidth]{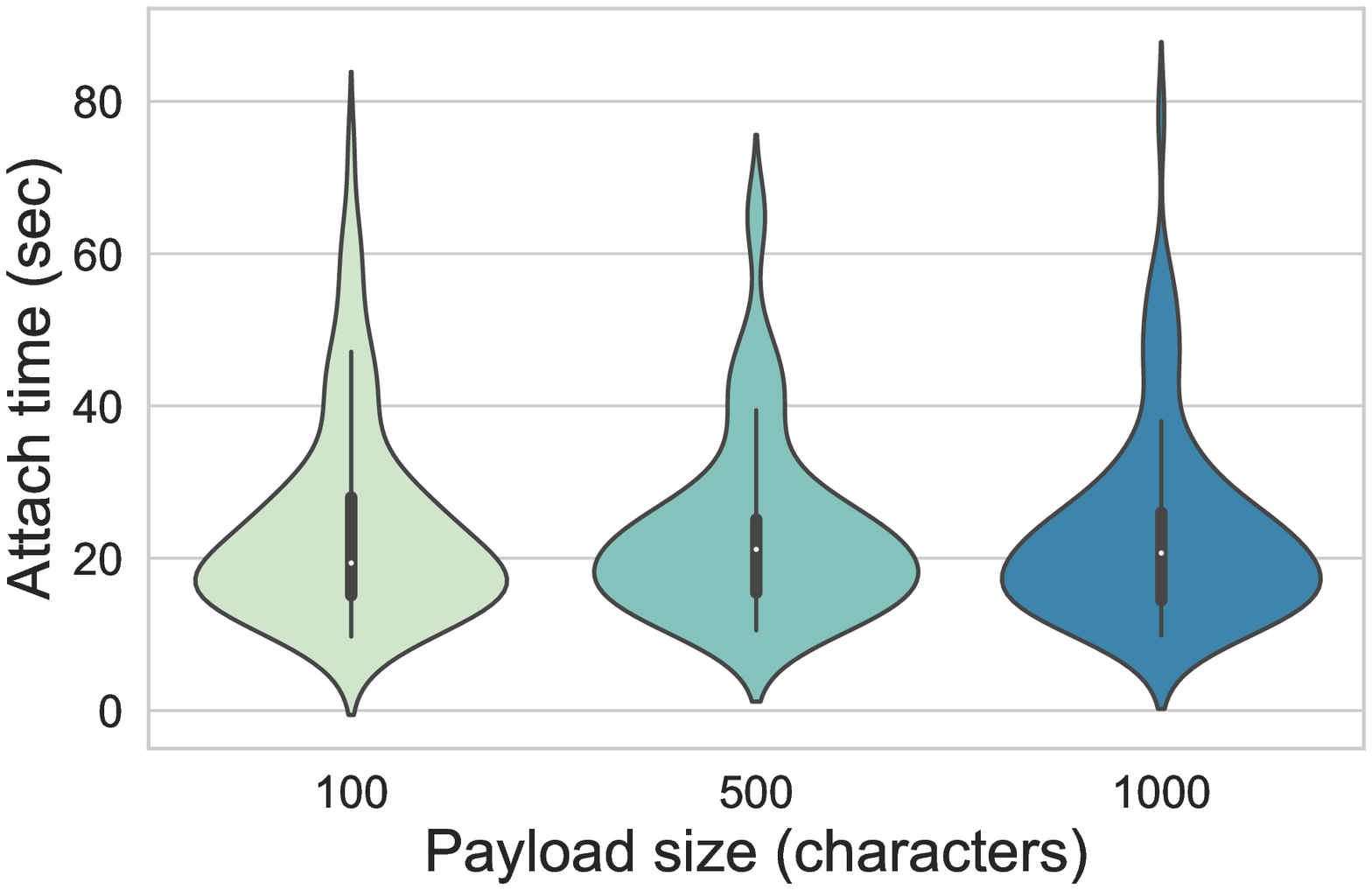}
\caption{}
\label{graph:RPIattach}
\end{subfigure}%
\hspace*{\fill}
\begin{subfigure}[b]{0.48\textwidth}
\centering
\includegraphics[width=\textwidth]{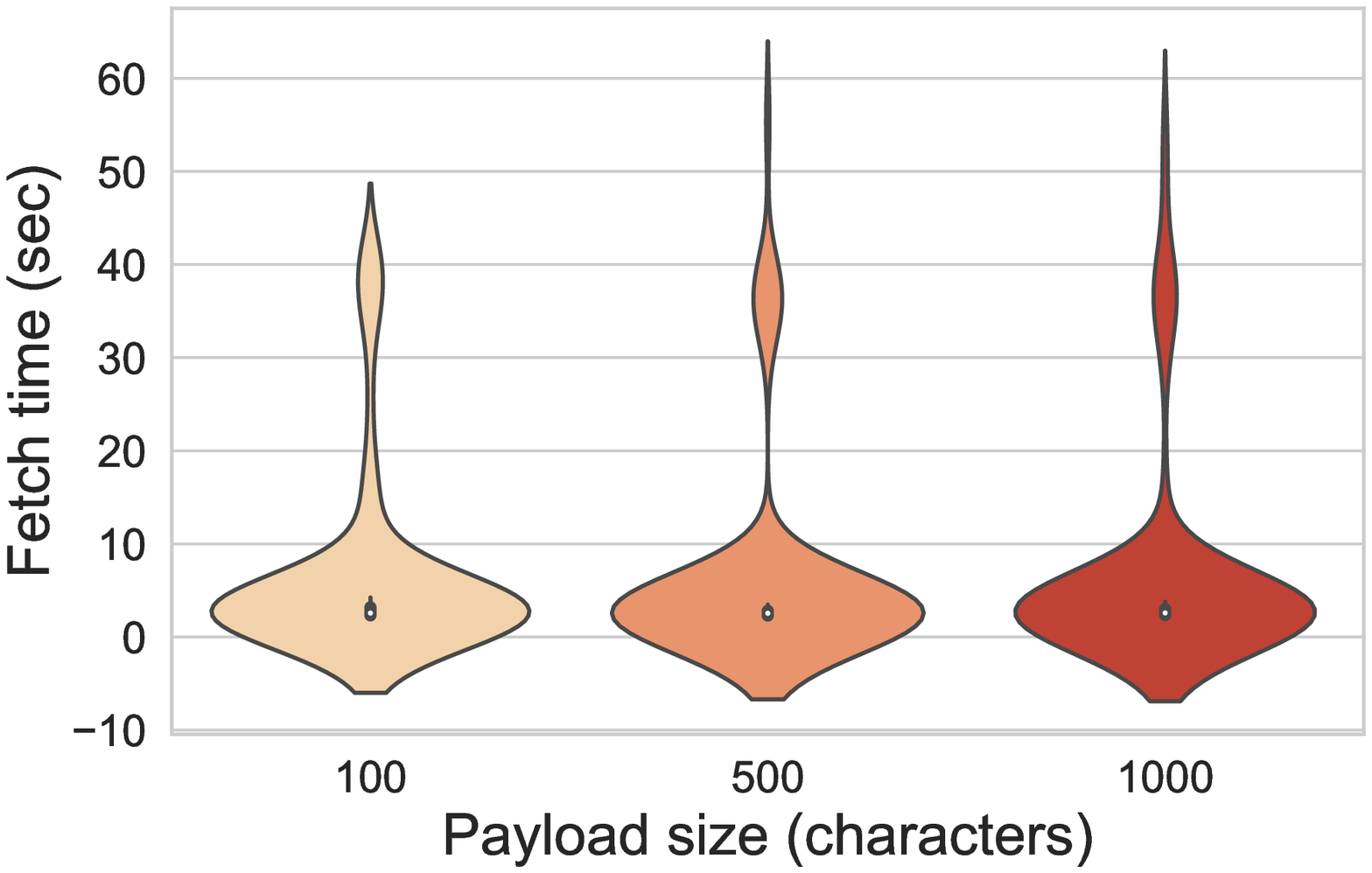}
\caption{}
\label{graph:RPIfetch}
\end{subfigure}
\centering
\caption{Raspberry Pi 3B: Estimated time (in sec) required to (a) attach data to the tangle, (b) fetch data from the tangle.}
\label{graph:RPI_attach_fetch}
\end{figure}

\begin{figure}[!ht]
\centerline{\includegraphics[width=3.0in]{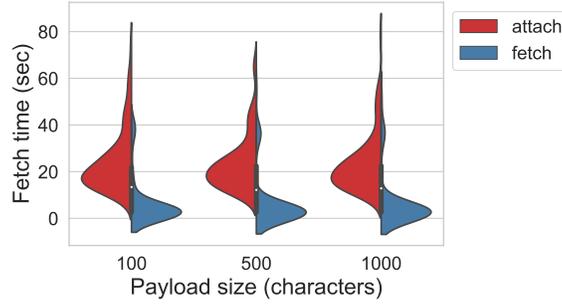}}
\caption{Comparison between estimated time (in sec) required by Raspberry Pi 3B to (a) attach data to the tangle, (b) fetch data from the tangle.}
\label{graph:splitRPI}
\end{figure}

Similarly, to simulate the proposed scheme on an IoT platform, we use Raspberry Pi 3B. On this platform, we consider attaching and fetching process in terms of time and energy constraints. In order to include sensor data, we use the Digital Humidity Temperature sensor (DHT11). We assume that any other sensor data can be integrated in a similar way to attach and fetch sensor data to/from the tangle, respectively.
Firstly, we compute the average time required to attach and fetch payload (including sensor data) to and from the tangle, respectively (shown in Fig.~\ref{graph:RPI_attach_fetch}). 
In particular, Fig.~\ref{graph:splitRPI} represents the distribution of data for performing IOTA operations remote such that the median value for attaching payload is around 20 sec while the median value for fetching data is around 2 sec.

\begin{figure}[!ht]
\centerline{\includegraphics[width=3.0in]{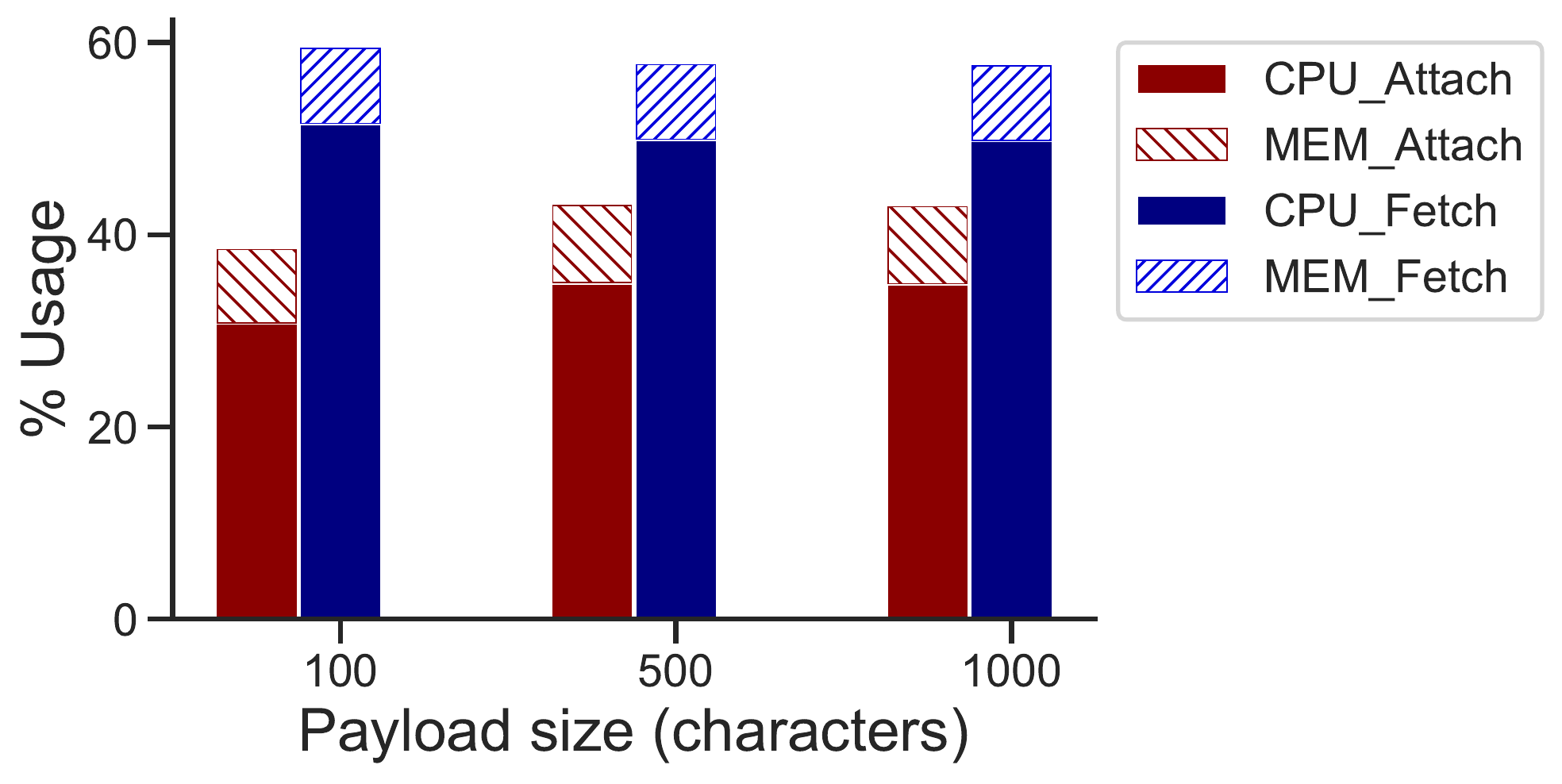}}
\caption{Comparison: CPU and memory usage during attaching and fetching data to the tangle. }
\label{graph:usage}
\end{figure}

Secondly, we compute the energy consumption by CPU. Out of 4 cores of Raspberry Pi 3B, a single core (power consumption= 221.0 mW per core) is utilized for attaching and fetching data. The measured average time and energy consumption (evaluated 100 times for security level 3) are given in Table~\ref{tab:energy}. It is important to mention that we only consider Raspberry Pi 3B as a \textit{light node} with the Minimum Weight Magnitude (MWM) parameter set to 14; where MWM is the PoW complexity currently used in the IOTA mainnnet. Thirdly, we compute the CPU and memory consumption during attaching and fetching phases. Irrespective of payload size, the fetching process consumes more CPU and memory (as shown in Fig.~\ref{graph:usage}). Since during fetching operation, tasks are carried out by Raspberry Pi 3B itself rather than the remote node, hence, it consumes more CPU and memory.

 \subsection{Informal Security Analysis} \label{securityAnalysis}
In this subsection, we revisit the security claims mentioned in Section~\ref{security_goals} and justify them to evaluate the performance of our proposed provenance-based scheme for ESC.

\textit{Claim 1: Data confidentiality}.

\textit{Justification}:
 The data is stored on the channel in encrypted form. Hence, only those $D_r$ having access to the $Channel_{ID}$ and authorization key ($\mathcal{K}$) can obtain and decrypt the payload. 

\begin{figure}[!ht]
\centerline{\includegraphics[width=3.0in]{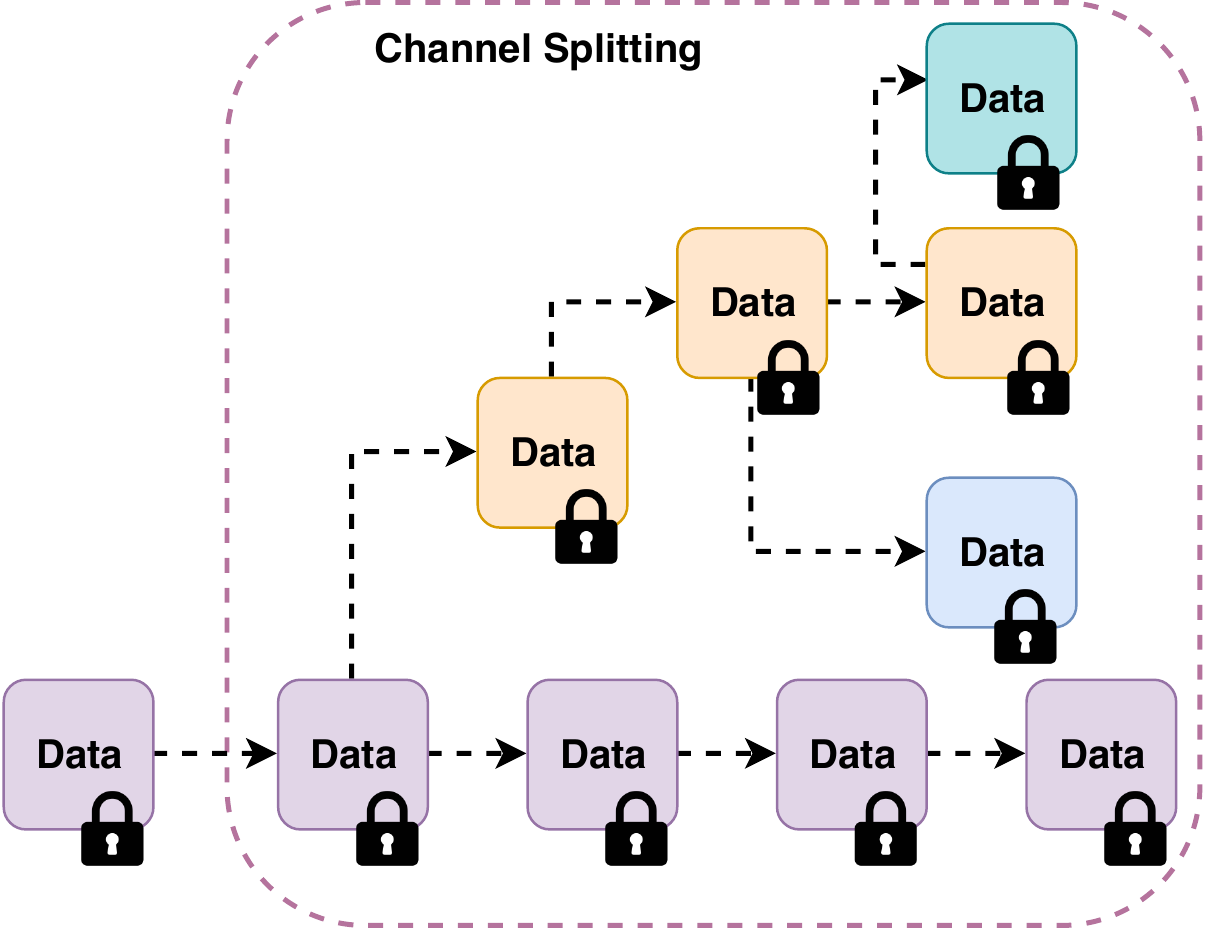}}
\caption{Channel splitting: sharing only a subset of data. }
\label{fig:channelsplit}
\end{figure}

\textit{Claim 2: Access control rights}.

\textit{Justification}:
MAM channel enables the off-shooting channel (as shown in Fig. \ref{fig:channelsplit}) particularly when the entirety of data is not intended to be shared. Such fine-grained access to data is desired in many scenarios in ESC. For example, the retailer may share the sales data or customer buying pattern data with the marketing companies while preserving the customer Personally Identifiable Information (PII). Similarly, the idea of channel splitting can also be used to limit access to a company's trade secrets from joint ventures, suppliers, distributors, or customers. Another significant use-case example scenario is when a company is buying some of its product's components from its competitor company. Defining access rights (i.e., grant and revoke) on data is based on an authorization key ($\mathcal{K}$) used in the restricted mode. The key is exchanged with the legitimate SC players only and can be changed to revoke access rights without any need to change the $Channel_{ID}$. The other modes provided by MAM channel includes \textit{public} and \textit{private}. Further details related to the MAM channel are provided in Section \ref{MAM}.  

To illustrate the process of fine-grained access through channel splitting, let us consider a scenario. Suppose, a seller $S_1$ (SID: SK\_GIH003) is selling components (for instance, DRAM chips) to a buyer $B_3$. $S_1$, also outsources its components to one of its partner sub-seller $S_{sub}$ (SID: CN\_SHE005) who further sells components to other buyers $B_1$ and $B_2$. $S_1$  and $S_{sub}$ define access control rights for their buyers so that they are able to retrieve the required information from them. The information, in particular, can be generalized as $T_{Data}$, $A_{Data}$, $Sales_{info}$ (showing sales pattern), $Client_{info}$ (list of clients), $Manufacturing_{info}$ (manufacturing process), $Advertising_{info}$ (advertising strategies). The defined policies and a few example queries are discussed in Table \ref{tab:channelspilit}. The query results can be retrieved on the basis of provenance key elements.

 \begin{table*}[ht!]
  \begin{tabular}{p{2.0cm}p{5.5cm}p{4.0cm}p{2.5cm} } 
  \hline
\textbf{Channel ID} & \textbf{Policies} & \textbf{Queries} & \textbf{Result} \\ [0.5ex] 
 \hline
 \hline
SID: SK\_GIH003  & Allow $B_3$ to access: 

$T_{Data}$, $A_{Data}$ & Fetch:

$Sales_{info}$ from $S_{1}$  &  Access Denied   \\ \\ \\

    & Allow $S_{sub}$ to access:
    
    $T_{Data}$, $A_{Data}$,
    
    $Sales_{info}$, $Client_{info}$, 
    
    $Advertising_{info}$  & 
    Fetch:
    
    $Manufacturing_{info}$from $S_{1}$  & Access Denied   \\ 
        &  & Fetch:
        
        $Sales_{info}$ from $S_{1}$  & Access Granted   \\ \\

 \hline
 
SID: CN\_SHE005 & Allow $B_1$ to access:

$T_{Data}$, $A_{Data}$, 

$Client_{info}$ & Fetch:

$Client_{info}$ from $S_{sub}$  & Access Granted  \\ \\ \\

    & Allow $B_2$ to access:
    
    $T_{Data}$, $A_{Data}$, 
    
    $Sales_{info}$ & Fetch:
    
    $Client_{info}$ from $S_{sub}$  & Access Denied   \\

\hline
\end{tabular}
\caption{Channel splitting example scenario: fine-grained access rights.} \label{tab:channelspilit}
\end{table*}

\begin{figure}[!ht]
\centerline{\includegraphics[width=4.0in]{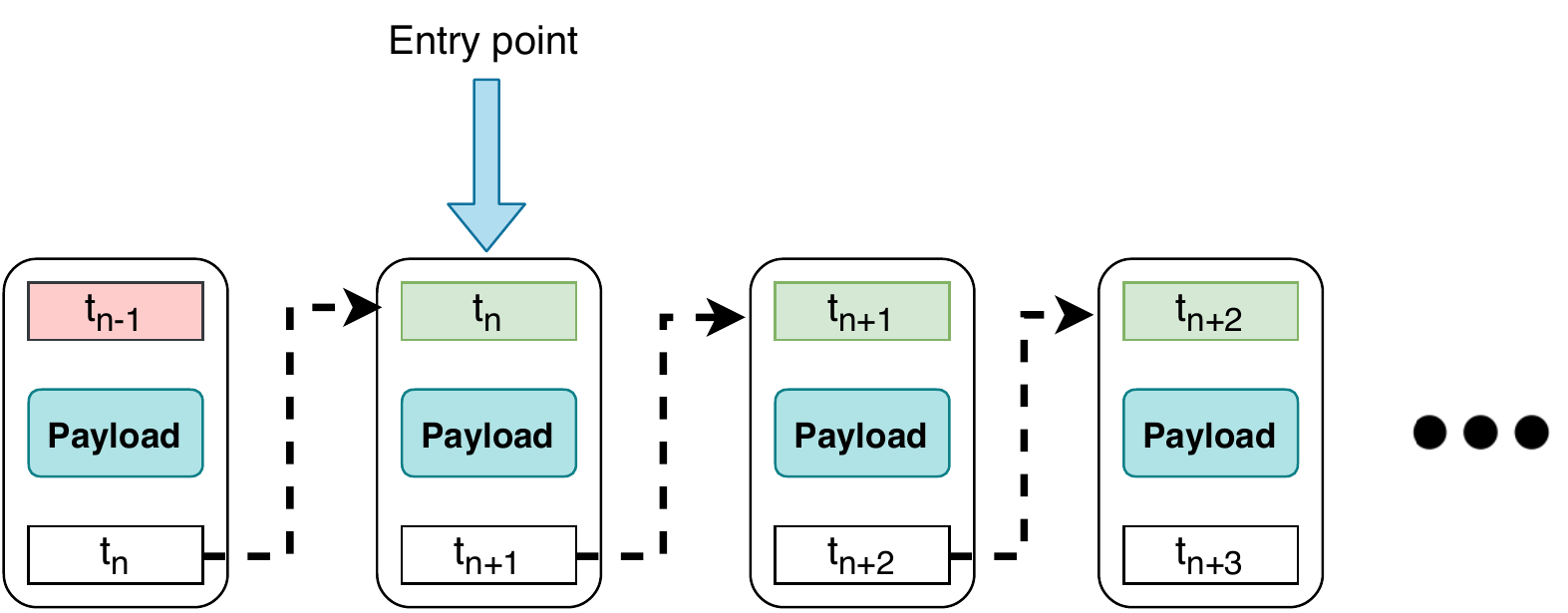}}
\caption{Message chain showing \emph{transaction linking} among transactions and \emph{forward secrecy} enforced at and after the point of entry. }
\label{fig:forwardlinking}
\end{figure}

\textit{Claim 3: Restrictions on data retrieval}.

\textit{Justification}:
To enforce a restriction on previous data in the message chain, we exploit forward secrecy, i.e., the subscriber can only locate and retrieve transactions at or after their point of entry in the channel, but not before their point of entry (as shown in Fig.~\ref{fig:forwardlinking}). Upon locating the transaction $t_n$, the subscriber can retrieve the address of the next transaction $t_{n+1}$ (as shown in Fig.~\ref{fig:forwardlinking}). 
When the masked message of one generation is decrypted, unmasked message contains {\it nextRoot} that is used for viewers to find the message of the next generation of the channel.

\textit{Claim 4: Preserving data integrity during trade events.} 

\textit{Justification}: 
Sensor tampering or data modification may occur during data creation or data transit phase thereby causing the GIGO problem. For example, in our SC case, the sensor can be tampered to change the readings leading to false data creation or the sensor data can be maliciously altered during communication. To address the former problem, we can rely on any of the existing solutions such as the use of Physical Unclonable Functions (PUF) provided in~\cite{javaid2018blockpro}. For the latter problem, we employed the MAM channel restricted mode for secure communication. Furthermore, it is assumed that IoT sensors are calibrated periodically. 
Another concern is how to ensure that a rogue participating entity is not corrupting the data? 
Some of the following solutions in the literature can be leveraged to solve this issue. A rating system for buyers and sellers based on previous trade events can be used. For instance, in \cite{khaqqi2018incorporating}, the authors used a reputation-based trading system which allows high reputations sellers to access better offers from the buyers. Another such solution is proposed in \cite{ramachandran2017using, brody2017blockchain} to enable monetary punishment mechanism to discourage any malicious changes in data or revoking trader's participation in the SC based on trust evaluation~\cite{malik2019trustchain}. 
Though all of the above solutions are proposed in combination with the blockchain ledger; however, they can be incorporated into the IOTA through oracles on the top (i.e., the concept of smart contracts in IOTA is underway~\cite{smartContractIOTA, smartContractIOTA2}).

To eliminate the GIGO problem in our proposed scheme, we adopted a combination of the below-mentioned solutions to mitigate the effect of adulteration. Firstly, we include product traceability through provenance data at each intermediate step in SC processes. Secondly, the inclusion of the warranty parameter can also help in establishing trust among entities in the long run. Thirdly, the verification of the transactions by both buyer and supplier can also confirm any fraudulent activity, for instance, $Receipt$ transaction in our proposed solution. Finally, the attacker can not repudiate once found guilty (claim 5).  
It is also worth noting that the adversary in a permissioned ledger requires consideration of many other possible security scenarios (for instance, attacks by an internal or external adversary, colluding users, etc.) and is out of the scope of this paper.

\textit{Claim 5: Non-repudiation}.

\textit{Justification}: 
To handle repudiation of trade events by any of the participating entities, transactions on the IOTA ledger ensure the immutability of data, the existence of SC events, and associated data carried out by the particular entity.

\subsection{Outstanding Issues and Challenges in IOTA Ledger} \label{discussionIOTA}

Analogous to the blockchain, IOTA also faces security and stability issues. Currently, IOTA is relying on a coordinator \textit{COO} for consensus that is responsible for the continuous generation of trust-able transactions to help to secure the infant tangle network from a double-spending attack. 
The two main problems arise due to the presence of COO include (i) single point of attack that can paralyze the whole tangle if COO stops working or taken over, and (ii) curtailing scalability of IOTA.
However, to optimize the designed system, \textit{Coo-less IRI} (CLIRI)~\cite{Coordicide} is recently introduced which is considered to be an important step towards the maturity of IOTA protocol, thereby making it a permissionless and scalable DLT. \textit{znet} is also launched as the first iteration of the coordinator-less testnet~\cite{COOless}. Hence, research is still ongoing on key improvement areas that would allow the desired decentralization. Keeping in mind the compatibility requirements for IOTA, we assume that the proposed scheme will be compatible with any up-gradation in the IOTA or the application layer MAM protocol.

Similar to the \emph{forking} problem in blockchain, tangle also suffers from \emph{parasite chains} in which an attacker makes a side tangle to double-spend the money. This problem can also be referred to as double-spending attack \cite{asokan1997state} discussed in detail in \cite{popov2017iota}. 
Though according to IOTA paper, parasite chain attack can be prevented when nodes use the MCMC tip selection strategy under the assumption that the main tangle has more hashing power than the attacker. However, as opposed to the assumption, attack analysis is still under discussion by the community \cite{parasite}. In the literature, \cite{cullen2019distributed} suggested a solution to simulate against parasite attack by proposing the matrix model for the MCMC tip selection algorithm.

For highly energy-constrained IoT devices (such as battery-powered) performing computationally expensive tasks is not practically possible without hardware-accelerated cryptography.
Nevertheless, powerful devices such as Raspberry Pi are still capable of doing IOTA operations as light node~\cite{elsts2018distributed}.

Another important issue is the staggering amount of transactions received by IOTA nodes which off course results in ever-increasing memory and CPU requirements. To combat this situation, a snapshot is performed to either reduce the size of the tangle or to reduce the burden on memory-constrained nodes.  A snapshot essentially throws away all the transaction history and resets the IOTA ledger to a list of all the addresses that have a nonzero balance. Therefore, such a global snapshot prunes the database to create room for newer transactions. 
It is hard for node owners with limited storage (IoT devices), to store full transaction history. To handle such situations, a local snapshot feature can be used that allows node owners to delete old transactions and keep their tangle database small. This option facilitates faster synchronization, lower resource requirements, and eliminates the need to wait for global snapshots~\cite{localsnapshot}.
For many scenarios, data needs to be stored for an extensive period of time. In order to deal with such use cases, the IOTA Foundation provides a permanode solution called Chronicles~\cite{permanode}. This solution, enables node owners to have unbounded storage of the tangle’s entire history and makes it accessible at scale. 
In the context of SC, both permanode solution and local snapshot can be used depending on the situation and requirements. For instance, in the case of resource-constrained devices, local snapshot can be adopted, however, in the case of the SC process, permanode solution can be adopted. Such features can be incorporated into our proposed work upon finalizing these features by IOTA Foundation.

In this proposed framework, we adopt a 2-tier approach to orchestrate provenance in the ESC. In the first tier, we collect SC data flowing across each SC participant, store securely in a distributed IOTA ledger, and manage data access rights. In the second tier, we construct provenance data to trace and track the product journey at each intermediate step in the SC cycle. Such a 2-tier approach provides an optimal strategy to carry out SC processes. For example, the first tier resolves the problems of fragmented data repositories and enables the SC participants to hide their trade secrets by defining data access rights while the second tier resolves the problems of identification of counterfeit products and helps in achieving customer's trust.

\section{Conclusion and Future Research Directions} \label{conclusion}
In this paper, we target two key issues, i.e., disparate data repositories, and untrustworthy data disseminating in the electronics supply chain space. To address these obstacles, we employ IOTA permissioned ledger that uses MAM protocol at the application layer to allow distributed and trustworthy data accessible to only authorized SC players. By exploiting the MAM channel, the data publishers share their data securely while allowing the authorized data receivers to access it. Therefore, it is feasible to construct the product provenance story through trustworthy sources and immutable data streams flowing across the global supply chain. In addition to the construction of secure provenance information, we also measure the energy consumption by the Raspberry Pi 3B IoT-based platform to ensure the suitability of the proposed scheme on resource-constrained devices.

For future work, we plan to survey other existing DLTs and compare them with IOTA in terms of performance (scalability, latency, and throughput) and device resource usage (CPU, memory, and energy consumption). Depending on the infrastructure, different types of sensors are deployed in different SC application areas. Therefore, we also want to port the proposed provenance-enabled SC system to other ARM-based devices to evaluate their compatibility across different platforms. As discussed in Subsection~\ref{securityAnalysis} (Claim 4), evaluating traders in SC based on the mechanism of trust scores to facilitate honest buyers and sellers is also part of our forthcoming research. From the perspective of SC management, introducing trade finance process to replace traditional finance procedures and risk management (for example, environmental risks)
are other potential extensions of our proposed scheme.    
Regarding the applicability of the proposed approach, SC data can be monetized to allow other business communities to learn and analyze the current industry trends and traits. This involves the integration of the current hyped technologies such as Artificial Intelligence (AI) and Machine Learning (ML). For this purpose, querying data can be customized based on access privileges or anonymization techniques. Such information can also help in studying forecasting future events to avoid inaccuracies and ultimately take possible measures against the bullwhip effect.  
Lastly, addressing challenges (such as real-time performance, coexistence, and interoperability) associated with IIoT, impacting the proposed SC system are among other interesting areas that are required to be explored.


\begin{thebibliography}{100}
\bibitem{brody2017blockchain}
P.~Brody, How blockchain is revolutionizing supply chain management, Ernst
  Young.
\end{thebibliography}

\end{document}